\documentclass[%
preprint,
 amsmath,amssymb,
 aps,
prb,
floatfix,
]{revtex4-2}

\usepackage{graphicx}
\usepackage{xcolor}
\usepackage{dcolumn}
\usepackage{bm}
\usepackage{placeins}

\usepackage{fancyhdr}
\begin{document}

\title{First-Principles Electronegativity Scale from the Atomic Mean Inner Potential}

\author{Jin-Cheng Zheng}
\email{jczheng@xmu.edu.cn}
\affiliation{ Department of Physics, Xiamen University, Xiamen, 361005, China}
\affiliation{Department of Physics, Xiamen University Malaysia, Sepang, 439000, Selangor, Malaysia}

\begin{abstract}
Electronegativity is a cornerstone of chemical intuition, essential for rationalizing bonding, reactivity, and material properties. However, prevailing scales remain empirically derived, often relying on parameterized models or composite physical quantities. In this work, we introduce a universal electronegativity scale founded on the atomic mean inner potential (AMIP), also known as the average Coulomb potential, a fundamental, quantum-mechanical property accessible through both first-principles computation and electron-scattering experiments. Our scale, denoted $\chi_{\mathrm{AMIP},p}$, is an analytic function of just three ground-state atomic descriptors and carries explicit physical units. It demonstrates excellent agreement with established scales and successfully classifies bonding types across 358 compounds, including adherence to the metalloid ``Si rule". Beyond replicating known trends, $\chi_{\mathrm{AMIP,1/2}}$ proves to be a powerful predictive tool, accurately determining Lewis acid strengths for over 14,000 coordination environments ($R^2=0.93$) and $\gamma$-ray annihilation spectral widths for 36 elements ($R^2=0.97$), outperforming previous methods. By linking electronegativity directly to a measurable quantum property, this work provides a unified and predictive descriptor for electronic structure and chemical behavior across the periodic table.
\end{abstract}




\maketitle

\section{Introduction}\label{sec1}

Electronegativity is a foundational concept in chemistry, crucial for describing and predicting trends in chemical bonding and reactivity across the periodic table~\cite{Pauling1932,Pauling1939,Mulliken1934,allred1958scale,Allen1989}. Introduced by Pauling in 1932~\cite{Pauling1932}, it quantifies an atom's ability to attract electrons within a chemical bond. Despite its utility, the original formulation lacks a direct physical basis and well-defined units.

This practical significance has motivated numerous alternative definitions. 
Mulliken grounded electronegativity in atomic properties, defining it as the average of an atom's ionization energy and electron affinity~\cite{Mulliken1934}. 
This concept was later connected to density functional theory (DFT) by Parr \textit{et al.}~\cite{Parr1978}, 
who equated it to the negative of the electronic chemical potential, $-\partial E / \partial N$. 
Other scales have related electronegativity to configuration energies~\cite{Allen1989}, effective nuclear charge, 
or atomic size~\cite{allred1958scale,li2006estimation,zheng2024universal}. 
While these developments have enriched the theoretical framework, 
many of them rely on composite empirical parameters or face challenges in reconciling theoretical predictions with observed chemical behavior. 
For example, scales based on ionization energies are inherently energy-dependent, 
whereas those based on valence or size are tied to the electron density distribution.

A physically rigorous definition of electronegativity should be directly connected to the electronic potential—the fundamental force responsible for electron attraction. We propose that the mean inner potential (MIP) is an ideal candidate for this purpose. The MIP represents the spatial average of the electrostatic potential within an atom, a property that intrinsically incorporates both the electron density and the atomic size~\cite{Bethe1928}. It is a measurable quantity, accessible through both first-principles electronic structure calculations and electron-scattering experiments.

Herein, we introduce a new electronegativity scale derived directly from the MIP. This approach is both simple and parameter-free, requiring only the principal quantum number as an additional input. By anchoring our scale in this single, well-defined physical property, we avoid the ambiguities and adjustable parameters that complicate existing scales. The result is a transparent and analytically derived electronegativity measure with a clear quantum-mechanical interpretation, offering a unified framework for understanding chemical bonding.

\section{Results}\label{sec2}

\subsection{Mean Inner Potential}

The mean inner potential (MIP), denoted $V_0$, is defined as the volume average of the electrostatic potential $V(\mathbf{r})$ within a finite crystal, with the potential referenced to zero at a point infinitely far from the crystal. This quantity arises from the non-uniform distribution of electrons within the crystal and can be measured experimentally using techniques such as electron holography~\cite{gabor1948new,gajdardziska1993accurate,saldin1994mean,li1999measurement,kruse2003determination}.

The foundational expression for the MIP of a unit cell with volume $\Omega$ was first derived by Bethe~\cite{Bethe1928} and is given by
\begin{equation}
V_0 = \frac{1}{\Omega} \int_\Omega V(\mathbf{r}) \, d^3 r .
\end{equation}
By decomposing the total potential $V(\mathbf{r})$ into a sum of atomic contributions, $\sum_i V_i(\mathbf{r})$, this expression can be rewritten as~\cite{anishchenko1966calculation,Becker1990,rez1994dirac,spence2013high}
\begin{equation}
V_0 = \frac{1}{\Omega} \sum_i \int_{\text{crystal}} V_i(\mathbf{r}) \, d^3 r .
\end{equation}
For a monatomic crystal, application of Poisson’s equation leads to a form that explicitly reveals the dependence of the MIP on the second moment of the total charge density (including contributions from all electrons, both core and valence), $\rho_{ti}(r)$~\cite{Bethe1928,rez1994dirac,spence2013high}:
\begin{equation}
V_0 = -\frac{2\pi}{3\Omega} \sum_i \int_0^\infty r^2 \rho_{ti}(r) \, d^3 r
= \frac{2\pi}{3\Omega} \sum_i Z_i \langle r_{ti}^2 \rangle .
\end{equation}

By taking the Fourier transform, both the atomic electron density and the atomic electrostatic potential can be expressed in reciprocal space $\mathbf{k}$~\cite{Peng1999,zheng2005sensitivity,Zheng2009}:
\begin{equation}
f^{(x)}(\mathbf{k}) = \int \rho(\mathbf{r}) \exp\!\left( 2\pi i \mathbf{k}\cdot\mathbf{r} \right)\, d^3 r ,
\label{eq:fx}
\end{equation}
\begin{equation}
f^{(e)}(\mathbf{k}) = \frac{1}{K} \int \phi(\mathbf{r}) \exp\!\left( 2\pi i \mathbf{k}\cdot\mathbf{r} \right)\, d^3 r .
\label{eq:fe}
\end{equation}
Here, $\mathbf{k}=2\mathbf{s}$ with $s=\sin\theta/\lambda$, and $K=h^2/(2\pi m_0 e)=47.87658$ when $\phi(\mathbf{r})$ is measured in electron volts~\cite{Peng1999}. The X-ray scattering factor $f^{(x)}(s)$ and the electron scattering factor $f^{(e)}(s)$ are related through Mott’s formula~\cite{mott1933}:
\begin{equation}
f^{(e)}(s) = \frac{A\,[Z - f^{(x)}(s)]}{s^2} .
\label{eq:mott}
\end{equation}
The coefficient $A=0.023934$ when $s$ is measured in units of \AA$^{-1}$. Here, $Z$ is the nuclear charge, $f^{(x)}(s)$ is given in units of the electron charge, and $f^{(e)}(s)$ has units of \AA.

We now introduce an important quantity, $f^{(e)}(0)$, known as the forward-scattering factor, or equivalently, the atomic scattering factor for fast electrons at zero scattering angle, corresponding to the limit $s \rightarrow 0$. For neutral atoms, the zero-angle electron scattering factor $f^{(e)}(0)$ is related to the second moment $\langle r_t^2 \rangle$ of the total charge density as
\begin{equation}
f^{(e)}(0) = \frac{a_0 \langle r_t^2 \rangle}{3} ,
\label{eq:fe0}
\end{equation}
where $a_0$ is the Bohr radius, $\langle r_t^2 \rangle$ is expressed in atomic units, and $f^{(e)}(0)$ is given in \AA.

When the electron scattering factor is parameterized as a sum of Gaussian functions,
\begin{equation}
f^{(e)}(s) = \sum_{i=1}^{N} a_i \exp\!\left( -b_i s^2 \right) ,
\label{eq:gaussian}
\end{equation}
as adopted in our previous work~\cite{Zheng2009}, the evaluation of $f^{(e)}(0)$ is straightforward:
\begin{equation}
f^{(e)}(0) = \sum_{i=1}^{N} a_i .
\label{eq:fe0sum}
\end{equation}

In a more general case, both $\langle r_t^2 \rangle$ and $f^{(e)}(0)$ can be obtained directly from all-electron atomic calculations based on density functional theory (DFT). Finally, the mean inner potential of a finite crystal can be expressed in terms of the forward-scattering factor $f^{(e)}(0)$ from electron scattering theory~\cite{rez1994dirac,Peng1999}:
\begin{equation}
V_0 = \frac{K}{\Omega} \sum_j n_j f_j^{(e)}(0),
\end{equation}
where $K = h^2/(2\pi m_0 e) = 47.87658$ when the scattering potential is measured in electron volts~\cite{Peng1999}, $n_j$ is the number of atoms of type $j$, and $f^{(e)}(0)$ is proportional to $\langle r_t^2 \rangle$. These atomic moments can be computed accurately using all-electron density functional theory~\cite{kohn1965self} or obtained from spectroscopic data or atomic scattering factor tables~\cite{rez1994dirac,Zheng2009}.

Interestingly, the above expressions demonstrate that the mean inner potential contains rich information about the atoms in a crystal, which motivates our proposal to construct a new electronegativity scale based on the MIP. As shown above, the MIP is directly related to the second moment of the charge density, indicating that it provides a measure of the effective ``size'' of an atom within a crystal and is therefore highly sensitive to the valence electron distribution~\cite{OKeeffe1994}.

Furthermore, Coulomb interactions play a central role in determining the structure and reactivity of molecules and solids. First, Coulomb interactions are essential for maintaining structural stability. In molecules, the attraction between positive and negative charges determines interatomic distances and bond angles, and thus the molecular geometry. In solids, the arrangement of positively charged ions and their interactions with electrons govern the stability and morphology of the crystal lattice. Second, Coulomb interactions also influence the reactivity of molecules and solids. Chemical reactions typically involve processes such as charge transfer and covalent bond formation, which are closely associated with Coulomb interactions. In solids, electron--electron Coulomb repulsion affects material properties such as electrical conductivity and optical response, thereby influencing the physical and chemical behavior of the material. Third, the Coulomb (electrostatic) potential is an essential component in describing the quantum behavior of atoms, molecules, and solids, as it constitutes a major part of the effective one-particle Hamiltonian in Hartree--Fock and Kohn--Sham formalisms~\cite{Becker1990}.
The expressions above establish that the MIP is fundamentally governed by the second moment of an atom’s charge density. It therefore provides a direct and robust measure of an atom’s spatial extent and its valence electron distribution. Since Coulomb interactions are central to chemical structure and reactivity, the MIP emerges as a well-grounded and physically transparent quantity on which to base a new electronegativity scale.

In the following, we describe our strategy for constructing an electronegativity scale based on the mean inner potential.

\subsection{Electronegativity from the Atomic Mean Inner Potential}

\begin{figure}[b]
  \centering
  \includegraphics[width=0.9\textwidth, angle=0]{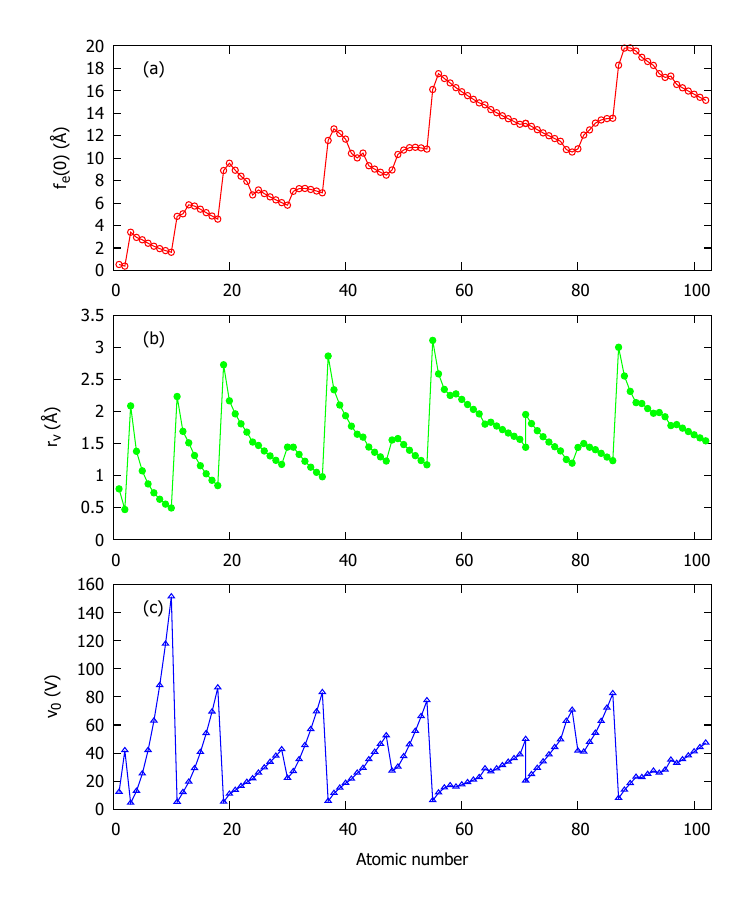}
  \caption{The forward electron scattering factor $f^{(e)}(0)$ (a), the valence radius $r_{v}$ (b), and atomic mean inner potential $v_{0}$ (c) for elements across the entire periodic table (from H (1) to No (102)).}
  \label{fig:Fig1} 
\end{figure}

\begin{figure}[ht!]
\centering
\includegraphics[width=0.70\linewidth, angle=-90]{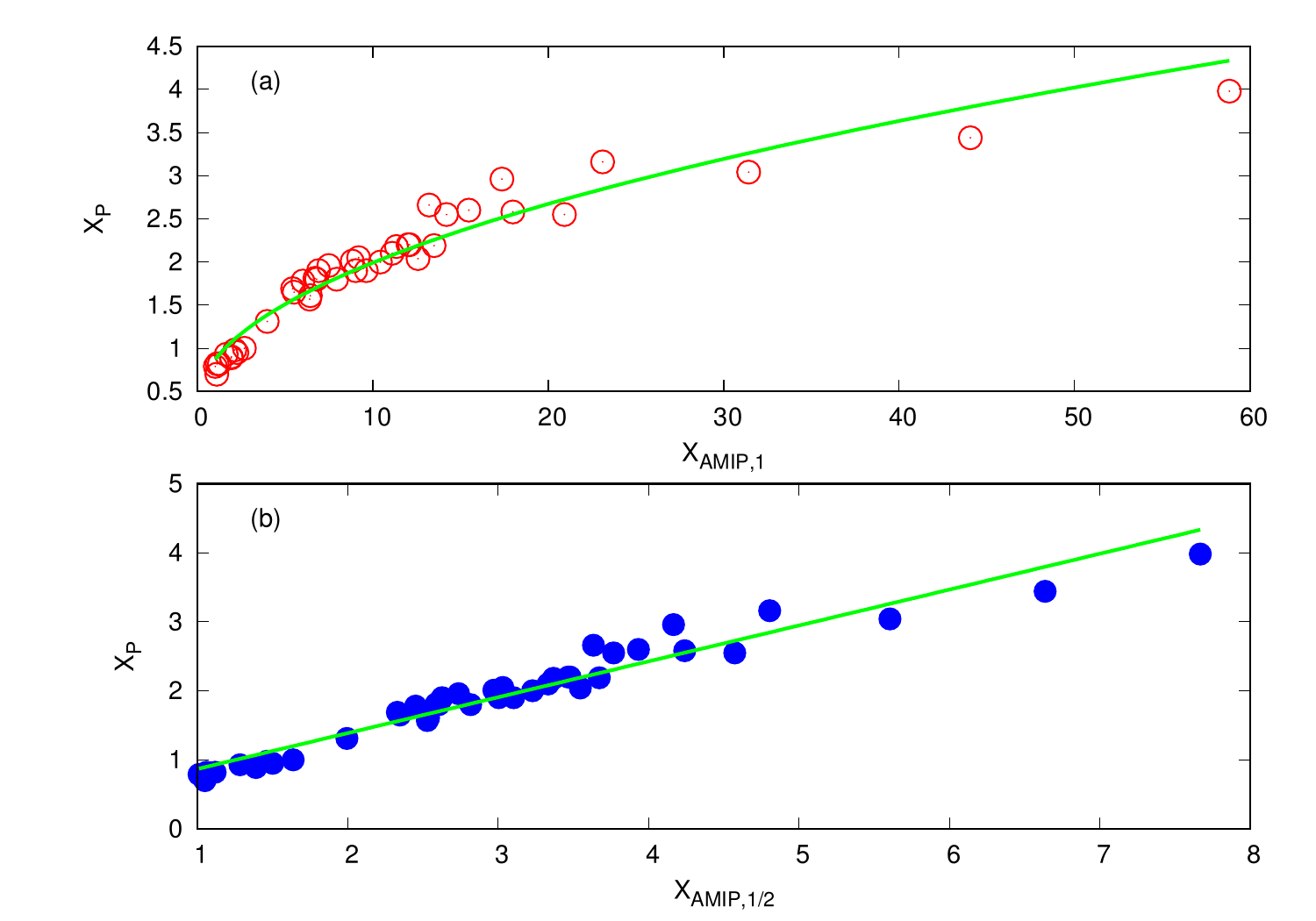}
\caption{Correlation between the proposed electronegativity scales and the Pauling scale ($\chi_P$) for main-group elements. (a) $\chi_P$ versus $\chi_{\mathrm{AMIP},1}$, fitted to a function of the form $a\sqrt{x}+c$, revealing a nonlinear relationship. (b) $\chi_P$ versus $\chi_{\mathrm{AMIP},1/2}$, fitted to a linear function $a x + c$ ($a=0.519$, $c=0.349$), demonstrating a strong linear correlation.}
\label{fig:Fig2}
\end{figure}

\begin{figure}
\centering
\includegraphics[width=1.00\linewidth]{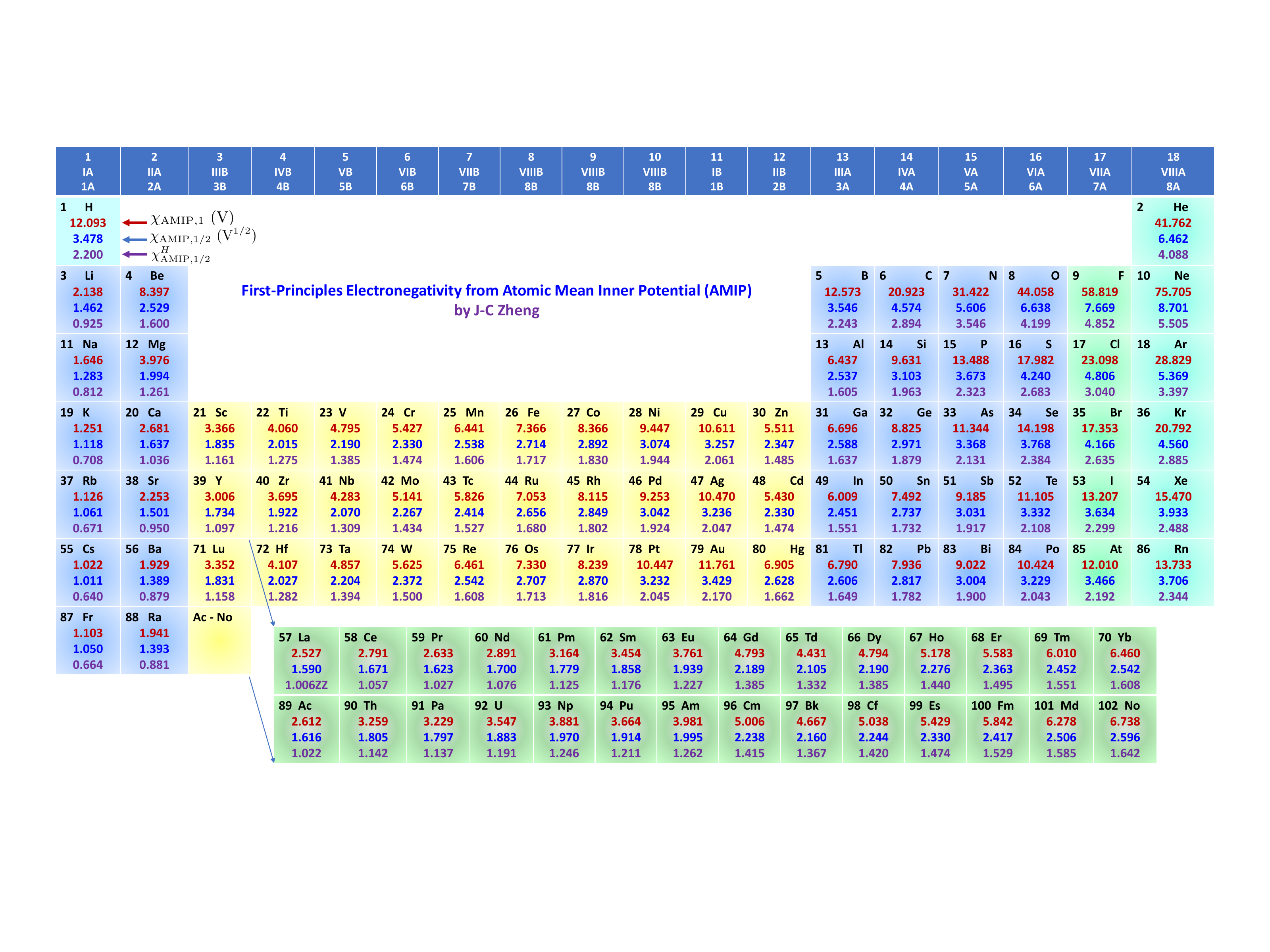}
\caption{Periodic table of elements (from H (1) to No (102)) showing the three proposed electronegativity scales, $\chi_{\mathrm{AMIP}}$ (V), $\chi_{\mathrm{AMIP},1/2}$ (V$^{1/2}$), and  $\chi_{\mathrm{AMIP},1/2}^H$ (dimensionless).} 
\label{fig:Fig3}
\end{figure}

We construct a new electronegativity scale based on the mean inner potential $V_0$. Since electronegativity is an intrinsic atomic property, the forward electron scattering factor $f^{(e)}(0)$ is, by definition, an atomic quantity, whereas the total crystal volume $\Omega$ is not. It is therefore necessary to define an atomic volume $\Omega_{a,j}$ in order to construct the atomic mean inner potential (AMIP), $v_{0j}$, for atom $j$ in the crystal:

\begin{equation}
v_{0j} = K \frac{f^{(e)}_j(0)}{\Omega_{a,j}} .
\label{eq:amip}
\end{equation}

Considering that $f^{(e)}(0)$ is directly related to the second moment of the total charge density, we use the mean valence atomic radius—defined as the first moment of the valence charge density—to determine the atomic volume. The mean valence atomic radius is given by
\begin{equation}
r_v = \frac{\int r\, \rho_v(\mathbf{r}) \, d^3 r}{\int \rho_v(\mathbf{r}) \, d^3 r} .
\label{eq:rv}
\end{equation}
For main-group elements with \textit{sp} valence configurations (e.g., $ns$ valence $s$ electrons and $np$ valence $p$ electrons), the mean valence atomic radius can be further expressed as
\begin{equation}
r_v (sp) = \frac{\int r \left[ n_s \rho_s(\mathbf{r}) + n_p \rho_p(\mathbf{r}) \right] d^3 r}{n_s + n_p}.
\label{eq:rvsp}
\end{equation}

More explicitly (e.g., within atomic DFT calculations), the mean valence atomic radius for main-group elements with \textit{sp} valence electrons can be expressed as
\begin{align}
r_v (sp) &= \frac{\int r \left[ n_s \rho_s(\mathbf{r}) + n_{p^{1/2}} \rho_{p^{1/2}}(\mathbf{r}) + n_{p^{3/2}} \rho_{p^{3/2}}(\mathbf{r}) \right] d^3 r}{n_s + n_{p^{1/2}} + n_{p^{3/2}}}  \\
    &= \frac{n_s r_s + n_{p^{1/2}} r_{p^{1/2}} + n_{p^{3/2}} r_{p^{3/2}}}{n_s + n_{p^{1/2}} + n_{p^{3/2}}} .
\label{eq:rvsp_explicit}
\end{align}

For transition metals with $3d$, $4d$, or $5d$ electrons, as well as lanthanides and actinides with $f$ electrons, the radial contraction effect arising from incomplete Slater shielding must be taken into account. As $d$ or $f$ electrons are progressively added across a transition-metal or lanthanide/actinide series, their screening of the nuclear charge is incomplete. In particular, $nd$ and $nf$ electrons are spatially localized and shield outer $ns$ electrons less effectively than $s$ electrons shield one another. As a result, the effective nuclear charge experienced by the valence $s$ orbital increases, leading to a systematic radial contraction that is most pronounced within a given principal quantum shell.

To account for this effect, we introduce a weighted, occupancy-based linear combination for transition metals and the lanthanide/actinide series:
\begin{align}
r_v (sdf)
&= \frac{\int r \left[ n_s \rho_s(\mathbf{r}) + \eta_d n_d \rho_d(\mathbf{r}) + \eta_f n_f \rho_f(\mathbf{r}) \right] d^3 r}{n_s + \eta_d n_d + \eta_f n_f}  \\
&= \frac{n_s r_s + \eta_d \left( n_{d^{3/2}} r_{d^{3/2}} + n_{d^{5/2}} r_{d^{5/2}} \right)
      + \eta_f \left( n_{f^{5/2}} r_{f^{5/2}} + n_{f^{7/2}} r_{f^{7/2}} \right)}
     {n_s + \eta_d n_d + \eta_f n_f} . \nonumber
\label{eq:rvdf}
\end{align}

Here, $\eta_d$ and $\eta_f$ denote the contraction factors for the transition-metal and lanthanide/actinide series, respectively. To minimize empirical parameters, we define $\eta_d$ and $\eta_f$ as the inverses of the maximum occupancies of the $d$ and $f$ orbitals, i.e., $\eta_d = 1/10$ and $\eta_f = 1/14$. 
This choice is physically motivated to account for orbital contraction effects while maintaining a parameter-minimal framework. Alternative contraction-correction schemes may also be considered, and their systematic evaluation is left for future work.

This formulation is fully consistent with the treatment of $sp$-orbital elements and is based on charge-density-weighted radial moments, thereby providing a rigorous theoretical foundation. The weighted linear combination ensures that the resulting valence radii properly reflect the spatial distribution of electrons across all relevant shells.

The corresponding mean valence atomic volume is then defined as
\begin{equation}
\Omega_{a,j} = \frac{4\pi}{3} r_{v,j}^3 .
\label{eq:omegaj}
\end{equation}
Substituting Eq.~\eqref{eq:omegaj} into Eq.~\eqref{eq:amip}, we obtain the AMIP as, 
\begin{equation}
v_{0j} = \frac{K f_j^{(e)}(0)}{\Omega_{a,j}} = \frac{K f^{(e)}_j(0)}{\frac{4\pi}{3} r_{v,j}^3} .
\label{eq:amipfinal}
\end{equation}

In this formulation, the dimension of $v_{0j}$ is the same as that of the mean inner potential $V_0$ (i.e., volts), as defined in Eq.~(11). The mean inner potential of a finite crystal can therefore be rewritten as
\begin{equation}
V_0 = \sum_j \left( \frac{n_j \Omega_{a,j}}{\Omega} \right) v_{0j} .
\label{eq:V0recast}
\end{equation}

The plots of three key quantities—the forward electron scattering factor $f^{(e)}(0)$ (a), the valence radius $r_v$ (b), and the atomic mean inner potential $v(0)$ (c)—for the main group elements are shown in Fig.~\ref{fig:Fig1}.

To construct an electronegativity scale, we normalize the atomic mean inner potential $v_0$ by the principal quantum number $n_q$, yielding
\begin{equation}
\label{eq:XAMIP-lin}
\chi_{\text{AMIP},1} = \frac{v_0}{n_q}
= K \frac{f^{(e)}(0)}{n_q \Omega_a} .
\end{equation}
For clarity, the element-independent prefactor is placed on the left-hand side, while all element-dependent quantities appear on the right-hand side.

While $\chi_{\text{AMIP},1}$ exhibits a non-linear correlation with the Pauling scale (Fig.~\ref{fig:Fig2}a), a strong linear relationship is achieved by taking the square root of this quantity (Fig.~\ref{fig:Fig2}b):
\begin{equation}
\label{eq:XAMIP-sqr}
\chi_{\text{AMIP},1/2} = \left(\frac{v_0}{n_q}\right)^{1/2}
= \left(K \frac{f^{(e)}(0)}{n_q \Omega_a}\right)^{1/2}.
\end{equation}
Both expressions are fully analytic, contain no empirical parameters, and each term possesses a clear physical interpretation.

These scales can be reformulated in terms of charge-density moments as:
\begin{equation}
\label{eq:XAMIP-lin-R}
\chi_{\text{AMIP},1} = \frac{K}{4\pi}
\frac{\langle r_t^2 \rangle}{n_q (r_v)^3},
\end{equation}
and
\begin{align}
\label{eq:XAMIP-sqr-R}
\chi_{\text{AMIP},1/2}
&= \sqrt{\frac{K}{4\pi}}
\sqrt{\frac{\langle r_t^2 \rangle}{n_q (r_v)^3}}
= \sqrt{\frac{3K}{4\pi}}
\sqrt{\frac{f^{(e)}(0)}{n_q (r_v)^3}}.
\end{align}

By employing all-electron density functional theory (DFT) calculations for atoms, the values of $f^{(e)}(0)$ and $r_v$ can be obtained, and thus $\chi_{\text{AMIP}}$ can be determined. 
In this work, we performed all-electron atomic calculations using the \textit{ld1.x} module of \textit{Quantum~Espresso}~\cite{QE-2009}. 
The code employs fully converged atomic basis sets, and relativistic effects are treated explicitly through the Dirac equation for heavy elements, including spin–orbit coupling and self-interaction corrections. The revised Perdew–Burke–Ernzerhof (revPBE) generalized gradient approximation (GGA) functional~\cite{zhang1998comment}, which has been shown to yield binding energies in closer agreement with experimental values, was adopted in this study.

Within our framework, the radial moments, forward electron scattering factor, and atomic mean inner potential—as well as the proposed electronegativity scales—can be derived analytically. For the hydrogen atom, we obtain
\begin{align}
    \langle r_t^2 \rangle(\mathrm{H}) &= 3a_0^2, \\
    r_v(\mathrm{H})    &= \frac{3a_0}{2}, \\
    f^{(e)}(0)(\mathrm{H}) &= a_0 .
\end{align}
Since $n_q(\mathrm{H}) = 1$, the corresponding electronegativity measures become
\begin{align}
\label{eq:XAMIP-H}
\chi_{\text{AMIP},1}(\mathrm{H}) &= v_0(\mathrm{H}) 
= K \frac{f^{(e)}(0)(\mathrm{H})}{\Omega_a(\mathrm{H})}
= \frac{2K}{9\pi a_0}, \\
\chi_{\text{AMIP},1/2}(\mathrm{H}) &= \sqrt{v_0(\mathrm{H})}
= \sqrt{K} \sqrt{\frac{f^{(e)}(0)(\mathrm{H})}{\Omega_a(\mathrm{H})}}
= \sqrt{K} \sqrt{\frac{2}{9\pi a_0}} ,
\end{align}
where $a_0$ denotes the Bohr radius.

Notably, the values of $r_v(\mathrm{H})$ and $f^{(e)}(0)(\mathrm{H})$ are invariant with respect to the choice of exchange–correlation functional, including the local density approximation (LDA)~\cite{perdew1981self}, the Perdew–Burke–Ernzerhof (PBE) generalized gradient approximation (GGA)~\cite{perdew1996generalized}, and revPBE~\cite{zhang1998comment}. Hydrogen therefore serves as a natural and robust reference point. On this basis, a dimensionless electronegativity scale normalized to the electronegativity of hydrogen can be constructed.

Using hydrogen as this reference, we define the final normalized AMIP-based electronegativity by scaling the raw AMIP electronegativity to reproduce the Pauling electronegativity of hydrogen. Specifically, the normalized electronegativity for element $X$ is defined as
\begin{equation}
\label{eq:XAMIP-norm-H-v0}
\chi^{\mathrm{H}}_{\text{AMIP,1/2}}(\mathrm{X})
= \chi_\mathrm{P}(\mathrm{H}) 
\frac{\chi_{\text{AMIP},1/2}(\mathrm{X})}{\chi_{\text{AMIP},1/2}(\mathrm{H})}
= \chi_\mathrm{P}(\mathrm{H})
\sqrt{\frac{v_{0}(\mathrm{X})/n_q(\mathrm{X})}{v_0(\mathrm{H})} },
\end{equation}
where $\chi_\mathrm{P}(\mathrm{H}) = 2.20$ is the Pauling electronegativity of hydrogen, $v_{0}(X)$ is the atomic mean inner potential of element $\mathrm{X}$, and $n_q(\mathrm{X})$ is the principal quantum number of its valence shell.

Substituting the explicit expressions for the atomic mean inner potential yields

\begin{align}
\label{eq:XAMIP-norm-H}
\chi^{\mathrm{H}}_{\text{AMIP,1/2}}(\mathrm{X})
&= \chi_\mathrm{P}(\mathrm{H})
\sqrt{\frac{f^{(e)}(0)(\mathrm{X})}{n_q(\mathrm{X})\,\Omega_{a}(\mathrm{X})}
\left( \frac{\Omega_a(\mathrm{H})}{f^{(e)}(0)(\mathrm{H})} \right)} \\
&= \chi_\mathrm{P}(\mathrm{H})
\sqrt{\frac{9a_0^2}{8}}
\sqrt{\frac{\langle r_t^2(\mathrm{X}) \rangle}{n_q(\mathrm{X})\, [r_v(\mathrm{X})]^3}} \nonumber \\
&= \chi_\mathrm{P}(\mathrm{H})
\sqrt{\frac{27a_0^2}{8}}
\sqrt{\frac{f^{(e)}(0)(\mathrm{X})}{n_q(\mathrm{X})\, [r_v(\mathrm{X})]^3}} \nonumber,
\end{align}
which depends solely on atomic quantities obtained from all-electron calculations, namely the second moment of the total charge density $\langle r_t^2 \rangle$ (or, equivalently, the forward electron scattering factor $f^{(e)}(0)$) and the first moment of the valence charge density $r_v$, together with the principal quantum number $n_q$. All parameters carrying units related to electron scattering (such as $K$) cancel out in the final expression.

By construction, this normalization ensures that hydrogen reproduces its Pauling electronegativity exactly and renders the scale dimensionless (or, equivalently, expressed in the same units as the Pauling scale), making it directly comparable with conventional electronegativity scales. Alternatively, $\chi_\mathrm{P}(\mathrm{H})$ may be replaced by the hydrogen electronegativity from any other scale to adopt a different normalization. At the same time, the present definition preserves a clear physical interpretation: the electronegativity reflects the strength of the Coulomb potential generated by an atom, normalized by its spatial extent and valence-shell energy scale.

Our proposed scales—$\chi_{\text{AMIP},1}$ (Eqs. \ref{eq:XAMIP-lin}, \ref{eq:XAMIP-lin-R}), $\chi_{\text{AMIP},1/2}$ (Eqs. \ref{eq:XAMIP-sqr}, \ref{eq:XAMIP-sqr-R}), and the dimensionless $\chi_{\text{AMIP},1/2}^{H}$ (Eq. \ref{eq:XAMIP-norm-H-v0}, \ref{eq:XAMIP-norm-H})—are visualized for elements across the entire periodic table (from 1 H to 102 No) (Fig. \ref{fig:Fig3}) and provided in full numerical detail in Table \ref{tab:AMIP-electronegativity}.

\begin{table*}[t]
\caption{
The $f^{(e)}(0)$ (\AA), mean valence radius $r_v$ (\AA), atomic mean inner potential $v_0$ (V), and electronegativity scales of main-group elements, obtained from mean inner potentials using all-electron revPBE-GGA atomic calculations. Here, $Z_{\mathrm{tot}}$ denotes the total atomic charge (i.e., the atomic number), $Z_v$ the number of valence electrons, and $n_q$ the principal quantum number. The electronegativity scales shown are $\chi_{\mathrm{AMIP},1}$, $\chi_{\mathrm{AMIP},1/2}$, and the normalized scaled electronegativity $\chi_{\mathrm{AMIP},1/2}^H$, which is dimensionless (Pauling's hydrogen value of 2.20 is used for \(\chi_P(H)\)). Note that, in units of \AA, $f^{(e)}(0) = \tfrac{1}{3} \langle r_t^2 \rangle$, and $f^{(e)}(0)(H) = a_0$ for hydrogen. The ground-state electronic configuration~\cite{CRC2014} is used for all elements, except for Pd, for which the $4d^{9}5s^{1}$ configuration is adopted in order to obtain a well-defined $s$-orbital radius. Only the $s$ electrons are considered for Zn, Cd, and Hg in the table. If the $nd^{10}$ electrons are included in the valence, the resulting value of $\chi_{\mathrm{AMIP}}$ would be higher, for Zn, $\chi_{\mathrm{AMIP},1} = 11.864$, $\chi_{\mathrm{AMIP},1/2} = 3.444$, and $\chi_{\mathrm{AMIP},1/2}^H = 2.179$; for Cd, $\chi_{\mathrm{AMIP},1} = 10.150$, $\chi_{\mathrm{AMIP},1/2} = 3.186$, and $\chi_{\mathrm{AMIP},1/2}^H = 2.015$; for Hg, $\chi_{\mathrm{AMIP},1} = 11.243$, $\chi_{\mathrm{AMIP},1/2} = 3.353$, and $\chi_{\mathrm{AMIP},1/2}^H = 2.121$. For the No (102) element, the $f^{14}s^2$ valence electron configuration is considered. If only the $s$ electrons are taken into account, the resulting $\chi_{\mathrm{AMIP}}$ will be smaller.
\label{tab:AMIP-electronegativity}}
\resizebox{\textwidth}{!}{%
\begin{tabular}{cccccccccc}
$Z_{\mathrm{tot}}$ & Element & $Z_v$ & $n_q$ & $f^{(e)}(0)$ (Å) & $r_v$ (Å) & 
$v_0$(V) & $\chi_{\mathrm{AMIP},1}$ (V) & $\chi_{\mathrm{AMIP},1/2}$ (V$^{1/2}$) &
$\chi_{\mathrm{AMIP},1/2}^H$ \\
\hline
1 & H & 1 & 1 & 0.529 & 0.794 & 12.093 & 12.093 & 3.478 & 2.200 \\
2 & He & 2 & 1 & 0.384 & 0.472 & 41.762 & 41.762 & 6.462 & 4.088 \\
3 & Li & 1 & 2 & 3.404 & 2.088 & 4.276 & 2.138 & 1.462 & 0.925 \\
4 & Be & 2 & 2 & 2.944 & 1.380 & 12.794 & 6.397 & 2.529 & 1.600 \\
5 & B & 3 & 2 & 2.731 & 1.075 & 25.146 & 12.573 & 3.546 & 2.243 \\
6 & C & 4 & 2 & 2.422 & 0.871 & 41.846 & 20.923 & 4.574 & 2.894 \\
7 & N & 5 & 2 & 2.157 & 0.732 & 62.844 & 31.422 & 5.606 & 3.546 \\
8 & O & 6 & 2 & 1.942 & 0.632 & 88.116 & 44.058 & 6.638 & 4.199 \\
9 & F & 7 & 2 & 1.766 & 0.556 & 117.638 & 58.819 & 7.669 & 4.852 \\
10 & Ne & 8 & 2 & 1.620 & 0.496 & 151.410 & 75.705 & 8.701 & 5.504 \\
11 & Na & 1 & 3 & 4.817 & 2.234 & 4.938 & 1.646 & 1.283 & 0.812 \\
12 & Mg & 2 & 3 & 5.042 & 1.691 & 11.928 & 3.976 & 1.994 & 1.261 \\
13 & Al & 3 & 3 & 5.844 & 1.512 & 19.311 & 6.437 & 2.537 & 1.605 \\
14 & Si & 4 & 3 & 5.735 & 1.314 & 28.893 & 9.631 & 3.103 & 1.963 \\
15 & P & 5 & 3 & 5.455 & 1.155 & 40.464 & 13.488 & 3.673 & 2.323 \\

\end{tabular}}
\end{table*}

\begin{table*}[t]
\resizebox{\textwidth}{!}{%
\begin{tabular}{cccccccccc}
$Z_{\mathrm{tot}}$ & Element & $Z_v$ & $n_q$ & $f^{(e)}(0)$ (Å) & $r_v$ (Å) & 
$v_0$(V) & $\chi_{\mathrm{AMIP},1}$ (V) & $\chi_{\mathrm{AMIP},1/2}$ (V$^{1/2}$) &
$\chi_{\mathrm{AMIP},1/2}^H$ \\
\hline
16 & S & 6 & 3 & 5.146 & 1.029 & 53.946 & 17.982 & 4.240 & 2.683 \\
17 & Cl & 7 & 3 & 4.848 & 0.928 & 69.294 & 23.098 & 4.806 & 3.040 \\
18 & Ar & 8 & 3 & 4.572 & 0.845 & 86.487 & 28.829 & 5.369 & 3.397 \\
19 & K & 1 & 4 & 8.899 & 2.729 & 5.004 & 1.251 & 1.118 & 0.707 \\
20 & Ca & 2 & 4 & 9.552 & 2.167 & 10.724 & 2.681 & 1.637 & 1.036 \\
21 & Sc & 3 & 4 & 8.928 & 1.964 & 13.464 & 3.366 & 1.835 & 1.161 \\
22 & Ti & 4 & 4 & 8.395 & 1.808 & 16.240 & 4.060 & 2.015 & 1.275 \\
23 & V & 5 & 4 & 7.935 & 1.679 & 19.180 & 4.795 & 2.190 & 1.385 \\
24 & Cr & 6 & 4 & 6.727 & 1.524 & 21.708 & 5.427 & 2.330 & 1.474 \\
25 & Mn & 7 & 4 & 7.171 & 1.471 & 25.764 & 6.441 & 2.538 & 1.606 \\
26 & Fe & 8 & 4 & 6.847 & 1.385 & 29.464 & 7.366 & 2.714 & 1.717 \\
27 & Co & 9 & 4 & 6.553 & 1.308 & 33.464 & 8.366 & 2.892 & 1.830 \\
28 & Ni & 10 & 4 & 6.286 & 1.239 & 37.788 & 9.447 & 3.074 & 1.944 \\
29 & Cu & 11 & 4 & 6.040 & 1.176 & 42.444 & 10.611 & 3.257 & 2.061 \\
30 & Zn & 2 & 4 & 5.814 & 1.445 & 22.044 & 5.511 & 2.347 & 1.485 \\
31 & Ga & 3 & 4 & 7.050 & 1.444 & 26.784 & 6.696 & 2.588 & 1.637 \\
32 & Ge & 4 & 4 & 7.284 & 1.331 & 35.300 & 8.825 & 2.971 & 1.879 \\
33 & As & 5 & 4 & 7.304 & 1.225 & 45.376 & 11.344 & 3.368 & 2.131 \\
34 & Se & 6 & 4 & 7.214 & 1.132 & 56.792 & 14.198 & 3.768 & 2.384 \\
35 & Br & 7 & 4 & 7.076 & 1.052 & 69.412 & 17.353 & 4.166 & 2.635 \\
36 & Kr & 8 & 4 & 6.916 & 0.983 & 83.168 & 20.792 & 4.560 & 2.885 \\
37 & Rb & 1 & 5 & 11.580 & 2.865 & 5.630 & 1.126 & 1.061 & 0.671 \\
38 & Sr & 2 & 5 & 12.608 & 2.339 & 11.265 & 2.253 & 1.501 & 0.950 \\
39 & Y & 3 & 5 & 12.195 & 2.101 & 15.030 & 3.006 & 1.734 & 1.097 \\
40 & Zr & 4 & 5 & 11.700 & 1.934 & 18.475 & 3.695 & 1.922 & 1.216 \\

\end{tabular}}
\end{table*}

\begin{table*}[t]
\resizebox{\textwidth}{!}{%
\begin{tabular}{cccccccccc}
$Z_{\mathrm{tot}}$ & Element & $Z_v$ & $n_q$ & $f^{(e)}(0)$ (Å) & $r_v$ (Å) & 
$v_0$(V) & $\chi_{\mathrm{AMIP},1}$ (V) & $\chi_{\mathrm{AMIP},1/2}$ (V$^{1/2}$) &
$\chi_{\mathrm{AMIP},1/2}^H$ \\
\hline
41 & Nb & 5 & 5 & 10.432 & 1.772 & 21.415 & 4.283 & 2.070 & 1.309 \\
42 & Mo & 6 & 5 & 10.024 & 1.646 & 25.705 & 5.141 & 2.267 & 1.434 \\
43 & Tc & 7 & 5 & 10.454 & 1.601 & 29.130 & 5.826 & 2.414 & 1.527 \\
44 & Ru & 8 & 5 & 9.329 & 1.446 & 35.265 & 7.053 & 2.656 & 1.680 \\
45 & Rh & 9 & 5 & 9.027 & 1.365 & 40.575 & 8.115 & 2.849 & 1.802 \\
46 & Pd & 10 & 5 & 8.748 & 1.293 & 46.265 & 9.253 & 3.042 & 1.924 \\
47 & Ag & 11 & 5 & 8.489 & 1.228 & 52.350 & 10.470 & 3.236 & 2.047 \\
48 & Cd & 2 & 5 & 8.951 & 1.556 & 27.150 & 5.430 & 2.330 & 1.474 \\
49 & In & 3 & 5 & 10.330 & 1.578 & 30.045 & 6.009 & 2.451 & 1.551 \\
50 & Sn & 4 & 5 & 10.725 & 1.485 & 37.460 & 7.492 & 2.737 & 1.732 \\
51 & Sb & 5 & 5 & 10.937 & 1.396 & 45.925 & 9.185 & 3.031 & 1.917 \\
52 & Te & 6 & 5 & 10.971 & 1.312 & 55.525 & 11.105 & 3.332 & 2.108 \\
53 & I & 7 & 5 & 10.916 & 1.236 & 66.035 & 13.207 & 3.634 & 2.299 \\
54 & Xe & 8 & 5 & 10.810 & 1.169 & 77.350 & 15.470 & 3.933 & 2.488 \\
55 & Cs & 1 & 6 & 16.121 & 3.109 & 6.132 & 1.022 & 1.011 & 0.640 \\
56 & Ba & 2 & 6 & 17.529 & 2.587 & 11.574 & 1.929 & 1.389 & 0.879 \\
57 & La & 3 & 6 & 17.102 & 2.345 & 15.162 & 2.527 & 1.590 & 1.006 \\
58 & Ce & 3 & 6 & 16.703 & 2.251 & 16.746 & 2.791 & 1.671 & 1.057 \\
59 & Pr & 2 & 6 & 16.283 & 2.275 & 15.798 & 2.633 & 1.623 & 1.027 \\
60 & Nd & 2 & 6 & 15.919 & 2.189 & 17.346 & 2.891 & 1.700 & 1.076 \\
61 & Pm & 2 & 6 & 15.573 & 2.109 & 18.984 & 3.164 & 1.779 & 1.125 \\
62 & Sm & 2 & 6 & 15.243 & 2.033 & 20.724 & 3.454 & 1.858 & 1.176 \\
63 & Eu & 2 & 6 & 14.926 & 1.963 & 22.566 & 3.761 & 1.939 & 1.227 \\
64 & Gd & 3 & 6 & 14.753 & 1.803 & 28.758 & 4.793 & 2.189 & 1.385 \\
65 & Tb & 2 & 6 & 14.330 & 1.833 & 26.586 & 4.431 & 2.105 & 1.332 \\

\end{tabular}}
\end{table*}

\begin{table*}[t]
\resizebox{\textwidth}{!}{%
\begin{tabular}{cccccccccc}
$Z_{\mathrm{tot}}$ & Element & $Z_v$ & $n_q$ & $f^{(e)}(0)$ (Å) & $r_v$ (Å) & 
$v_0$(V) & $\chi_{\mathrm{AMIP},1}$ (V) & $\chi_{\mathrm{AMIP},1/2}$ (V$^{1/2}$) &
$\chi_{\mathrm{AMIP},1/2}^H$ \\
\hline
66 & Dy & 2 & 6 & 14.050 & 1.774 & 28.764 & 4.794 & 2.190 & 1.385 \\
67 & Ho & 2 & 6 & 13.779 & 1.718 & 31.068 & 5.178 & 2.276 & 1.440 \\
68 & Er & 2 & 6 & 13.517 & 1.665 & 33.498 & 5.583 & 2.363 & 1.495 \\
69 & Tm & 2 & 6 & 13.264 & 1.614 & 36.060 & 6.010 & 2.452 & 1.551 \\
70 & Yb & 2 & 6 & 13.019 & 1.566 & 38.760 & 6.460 & 2.542 & 1.608 \\
71 & Lu & 3 & 6 & 13.100 & 1.953 & 20.112 & 3.352 & 1.831 & 1.158 \\
72 & Hf & 4 & 6 & 12.834 & 1.812 & 24.642 & 4.107 & 2.027 & 1.282 \\
73 & Ta & 5 & 6 & 12.540 & 1.701 & 29.142 & 4.857 & 2.204 & 1.394 \\
74 & W & 6 & 6 & 12.253 & 1.607 & 33.750 & 5.625 & 2.372 & 1.500 \\
75 & Re & 7 & 6 & 11.998 & 1.524 & 38.766 & 6.461 & 2.542 & 1.608 \\
76 & Os & 8 & 6 & 11.748 & 1.451 & 43.980 & 7.330 & 2.707 & 1.713 \\
77 & Ir & 9 & 6 & 11.506 & 1.386 & 49.434 & 8.239 & 2.870 & 1.816 \\
78 & Pt & 10 & 6 & 10.769 & 1.252 & 62.682 & 10.447 & 3.232 & 2.045 \\
79 & Au & 11 & 6 & 10.549 & 1.195 & 70.566 & 11.761 & 3.429 & 2.170 \\
80 & Hg & 2 & 6 & 10.833 & 1.440 & 41.430 & 6.905 & 2.628 & 1.662 \\
81 & Tl & 3 & 6 & 12.060 & 1.501 & 40.740 & 6.790 & 2.606 & 1.648 \\
82 & Pb & 4 & 6 & 12.518 & 1.443 & 47.616 & 7.936 & 2.817 & 1.782 \\
83 & Bi & 5 & 6 & 13.127 & 1.405 & 54.132 & 9.022 & 3.004 & 1.900 \\
84 & Po & 6 & 6 & 13.407 & 1.348 & 62.544 & 10.424 & 3.229 & 2.042 \\
85 & At & 7 & 6 & 13.527 & 1.290 & 72.060 & 12.010 & 3.466 & 2.192 \\
86 & Rn & 8 & 6 & 13.558 & 1.234 & 82.398 & 13.733 & 3.706 & 2.344 \\
87 & Fr & 1 & 7 & 18.281 & 3.002 & 7.721 & 1.103 & 1.050 & 0.664 \\
88 & Ra & 2 & 7 & 19.810 & 2.554 & 13.587 & 1.941 & 1.393 & 0.881 \\
89 & Ac & 3 & 7 & 19.826 & 2.314 & 18.284 & 2.612 & 1.616 & 1.022 \\
90 & Th & 4 & 7 & 19.545 & 2.139 & 22.813 & 3.259 & 1.805 & 1.142 \\

\end{tabular}}
\end{table*}

\begin{table*}[t]
\resizebox{\textwidth}{!}{%
\begin{tabular}{cccccccccc}
$Z_{\mathrm{tot}}$ & Element & $Z_v$ & $n_q$ & $f^{(e)}(0)$ (Å) & $r_v$ (Å) & 
$v_0$(V) & $\chi_{\mathrm{AMIP},1}$ (V) & $\chi_{\mathrm{AMIP},1/2}$ (V$^{1/2}$) &
$\chi_{\mathrm{AMIP},1/2}^H$ \\
\hline
91 & Pa & 3 & 7 & 18.990 & 2.125 & 22.603 & 3.229 & 1.797 & 1.137 \\
92 & U & 3 & 7 & 18.616 & 2.046 & 24.829 & 3.547 & 1.883 & 1.191 \\
93 & Np & 3 & 7 & 18.265 & 1.973 & 27.167 & 3.881 & 1.970 & 1.246 \\
94 & Pu & 2 & 7 & 17.528 & 1.984 & 25.648 & 3.664 & 1.914 & 1.211 \\
95 & Am & 2 & 7 & 17.197 & 1.918 & 27.867 & 3.981 & 1.995 & 1.262 \\
96 & Cm & 3 & 7 & 17.315 & 1.781 & 35.042 & 5.006 & 2.238 & 1.415 \\
97 & Bk & 2 & 7 & 16.570 & 1.796 & 32.669 & 4.667 & 2.160 & 1.367 \\
98 & Cf & 2 & 7 & 16.271 & 1.741 & 35.266 & 5.038 & 2.244 & 1.420 \\
99 & Es & 2 & 7 & 15.981 & 1.688 & 38.003 & 5.429 & 2.330 & 1.474 \\
100 & Fm & 2 & 7 & 15.699 & 1.637 & 40.894 & 5.842 & 2.417 & 1.529 \\
101 & Md & 2 & 7 & 15.424 & 1.589 & 43.946 & 6.278 & 2.506 & 1.585 \\
102 & No & 2 & 7 & 15.157 & 1.543 & 47.166 & 6.738 & 2.596 & 1.642 \\
\end{tabular}}
\end{table*}

\subsection{Comparison with other electronegativity scales}

\begin{figure}[htbp]
\centering
\includegraphics[width=0.70\linewidth,angle=-90]{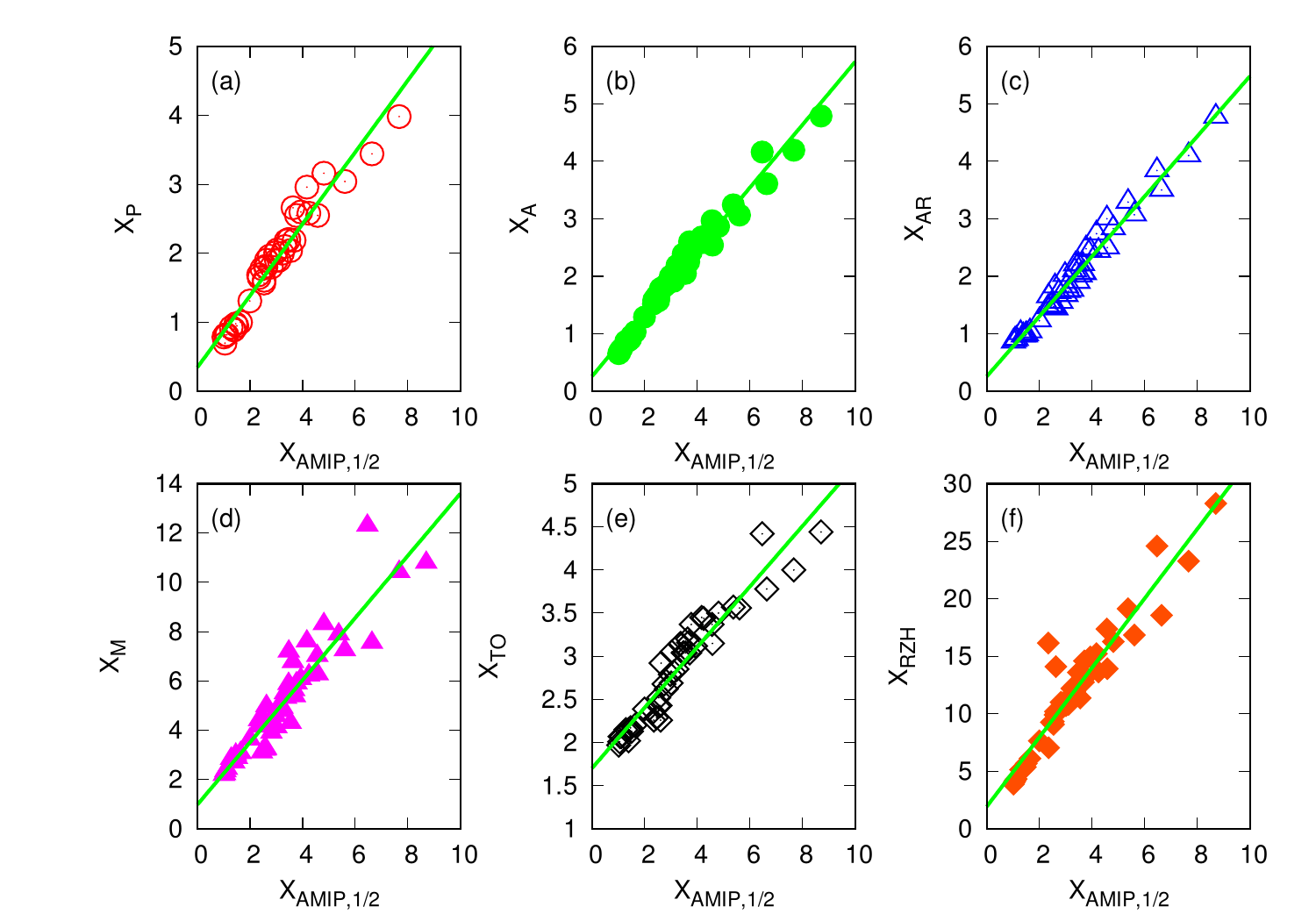}
\caption{
Parity plots comparing the proposed $\chi_{\text{AMIP},1/2}$ scale with various established electronegativity scales for main-group elements.
The fitting results are listed below: 
$\chi_{\text{P}}$ = 0.519382$\chi_{\text{AMIP},1/2}$ + 0.349086, 
$\chi_{\text{A}}$ = 0.546745$\chi_{\text{AMIP},1/2}$ + 0.262706, 
$\chi_{\text{AR}}$ = 0.521839$\chi_{\text{AMIP},1/2}$ + 0.261769, 
$\chi_{\text{M}}$ = 1.2617$\chi_{\text{AMIP},1/2}$ + 0.985428, 
$\chi_{\text{TO}}$ = 0.350907$\chi_{\text{AMIP},1/2}$ + 1.70328, 
$\chi_{\text{RZH}}$ = 3.0165$\chi_{\text{AMIP},1/2}$ + 1.94308, 
}
\label{fig:Fig4}
\end{figure}

\begin{figure}[htbp]
  \centering
  \includegraphics[width=0.9\textwidth, angle=0]{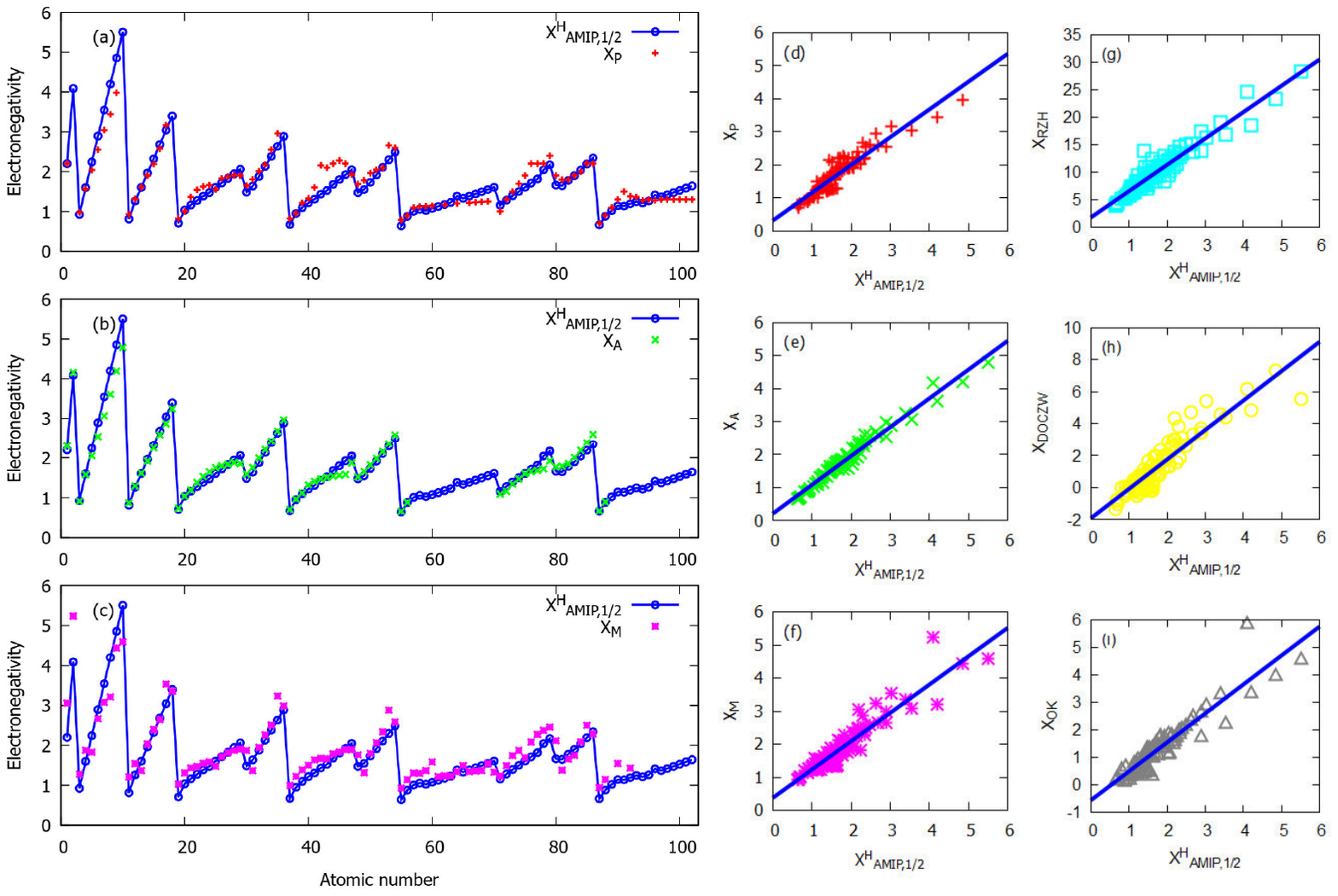}
\caption{Comparison of the proposed electronegativity scale $\chi^{\mathrm{H}}_{\mathrm{AMIP},1/2}$ with classical electronegativity scales: (a) Pauling, (b) Allen, and (c) Mulliken, for elements across the entire periodic table (from H (1) to No (102)). The coefficients of determination ($R^2$) between $\chi^{\mathrm{H}}_{\mathrm{AMIP},1/2}$ and the Pauling, Allen, and Mulliken scales are 0.8836, 0.9670, and 0.8732, respectively, as shown in panels (d)–(f). Note that the Mulliken scale has been rescaled to Pauling units. 
The $R^2$ values between $\chi^{\mathrm{H}}_{\mathrm{AMIP},1/2}$ and more recent electronegativity scales are also shown: (g) Rahm–Zeng–Hoffmann (RZH), 0.9169; (h) Dong–Oganov–Cui–Zhou–Wang (DOCZW), 0.8402; and (i) Oganov–Kostenko (OK), 0.8581.}
  \label{fig:Fig5_XAMIP12H_vs_XP-XA-XM} 
\end{figure}

The proposed scale, $\chi_{\text{AMIP},1/2}$ (Eqs.~\eqref{eq:XAMIP-sqr} and~\eqref{eq:XAMIP-sqr-R}), for main-group elements shows excellent agreement with several established electronegativity scales, including those of Pauling~\cite{Pauling1932,Pauling1939}, Allen~\cite{Allen1989,Allen1994}, Allred–Rochow~\cite{allred1958scale}, and Mulliken~\cite{Mulliken1934}, as well as the thermochemical scale of Tantardini and Oganov (TO)~\cite{tantardini2021thermochemical} and the valence binding-energy scale of Rahm, Zeng, and Hoffmann (RZH)~\cite{rahm2018electronegativity}. Parity plots illustrating these correlations are shown in Fig.~\ref{fig:Fig4}.

We also compared our proposed electronegativity scale with the classical scales of Pauling, Allen, and Mulliken, as well as recent scales such as RZH~\cite{rahm2018electronegativity}, Dong–Oganov–Cui–Zhou–Wang (DOCZW)~\cite{dong2022electronegativity}, and Oganov–Kostenko (OK)~\cite{oganov2025simple}, for the entire periodic table (from H (1) to No (102)), as shown in Fig.~\ref{fig:Fig5_XAMIP12H_vs_XP-XA-XM}.

\clearpage
The strength of these correlations was quantified using the coefficient of determination ($R^{2}$) for both 
main-group elements only and the entire periodic table, with detailed values provided in Table~\ref{tab:2}. 
These results indicate that $\chi_{\text{AMIP},1/2}$ correlates most strongly with the Allen and 
Allred--Rochow scales, followed closely by Pauling's classical thermochemical scale. 
It also maintains robust agreement with the Mulliken scale, as well as with the more recent TO, RZH, DOCZW, 
and OK scales. 

The somewhat lower correlation of our scale with the Mulliken scale can be attributed to the intrinsic dependence of the Mulliken electronegativity on experimental or calculated ionization energies and electron affinities. Both quantities can exhibit significant variations due to electronic relaxation, spin polarization, and finite basis set effects, especially for elements with partially filled d or f shells. These variations introduce deviations that reduce the linear correlation with the AMIP-based electronegativity scale. 

Each reference electronegativity scale is grounded in a distinct physical principle. Pauling’s definition is based on bond dissociation energies and reflects the tendency of an atom to attract electrons within a chemical bond~\cite{Pauling1932}. Allen’s scale is derived from the average valence-shell ionization energy, emphasizing intrinsic atomic properties independent of bonding environment~\cite{Allen1989}. The Allred–Rochow scale relates electronegativity to the effective nuclear charge experienced by valence electrons at the covalent radius, thereby linking electronegativity to electrostatic attraction~\cite{allred1958scale}. Mulliken’s definition is based on the arithmetic mean of the ionization energy and electron affinity, providing a direct energetic measure of an atom’s electron-accepting ability~\cite{Mulliken1934}.

The thermochemical (TO) scale~\cite{tantardini2021thermochemical} refines Pauling’s original approach by constructing a comprehensive, dimensionless electronegativity set for all elements, optimized to improve predictions of bond polarity and formation enthalpies across a wide range of compounds. In contrast, the RZH scale~\cite{rahm2018electronegativity} defines electronegativity as the ground-state average valence electron binding energy, yielding values that are conceptually related to, but physically distinct from, Allen’s ionization-based definition.

More recently, Dong \textit{et al.}  (DOCZW)~\cite{dong2022electronegativity} extended the Mulliken framework to high-pressure conditions by explicitly accounting for pressure-induced changes in ionization energies and electron affinities, thereby enabling pressure-dependent electronegativity and chemical hardness descriptors. Oganov and Kostenko (OK)~\cite{oganov2025simple} proposed an electronegativity-based model derived from large-scale formation enthalpy data, in which optimized elemental electronegativities serve as effective parameters for predicting compound stability across the periodic table. Together, these modern approaches highlight the continued evolution of electronegativity concepts toward first-principles, environment-aware, and data-informed descriptors.

The strong correlations (with all $R^{2} > 0.80$) demonstrate that the atomic mean inner potential (AMIP) captures a fundamental aspect of chemical bonding, effectively bridging thermochemical, spectroscopic, and electronic-structure perspectives. The consistently high $R^{2}$ values further underscore the predictive power and reliability of $\chi_{\text{AMIP},1/2}$ as a first-principles descriptor of chemical reactivity and bonding trends across the periodic table.

\begin{table}[htb]
    \centering
    \caption{Coefficient of determination ($R^{2}$) for the correlations between the proposed $\chi_{\mathrm{AMIP},1/2}$ scale and various established electronegativity scales, evaluated for main-group (MG) elements and for all elements across the periodic table (All), including transition metals as well as lanthanides and actinides. Note that lanthanides and actinides are not included in Allen's electronegativity scale.}
    \label{tab:2}
    \begin{ruledtabular}
        \begin{tabular}{ccc}
\textrm{Electronegativity scale} & \textrm{\(R^{2}(\text{MG})\)} & \textrm{\(R^{2}(\text{All})\)}\\
            \hline
Pauling                    & 0.9448    & 0.8836 \\
Allen                      & 0.9742    & 0.9670 \\
Allred--Rochow             & 0.9714    & NA \\
Mulliken                   & 0.8727    & 0.8732 \\
Tantardini--Oganov (TO)    & 0.9075    & 0.8032 \\
Rahm--Zeng--Hoffmann (RZH) & 0.9062    & 0.9169 \\
\end{tabular}
\end{ruledtabular}
\end{table}

\section{Applications}

\subsection{Classification of Metals and Nonmetals}

\begin{figure}
  \centering
  \includegraphics[width=0.80\linewidth]{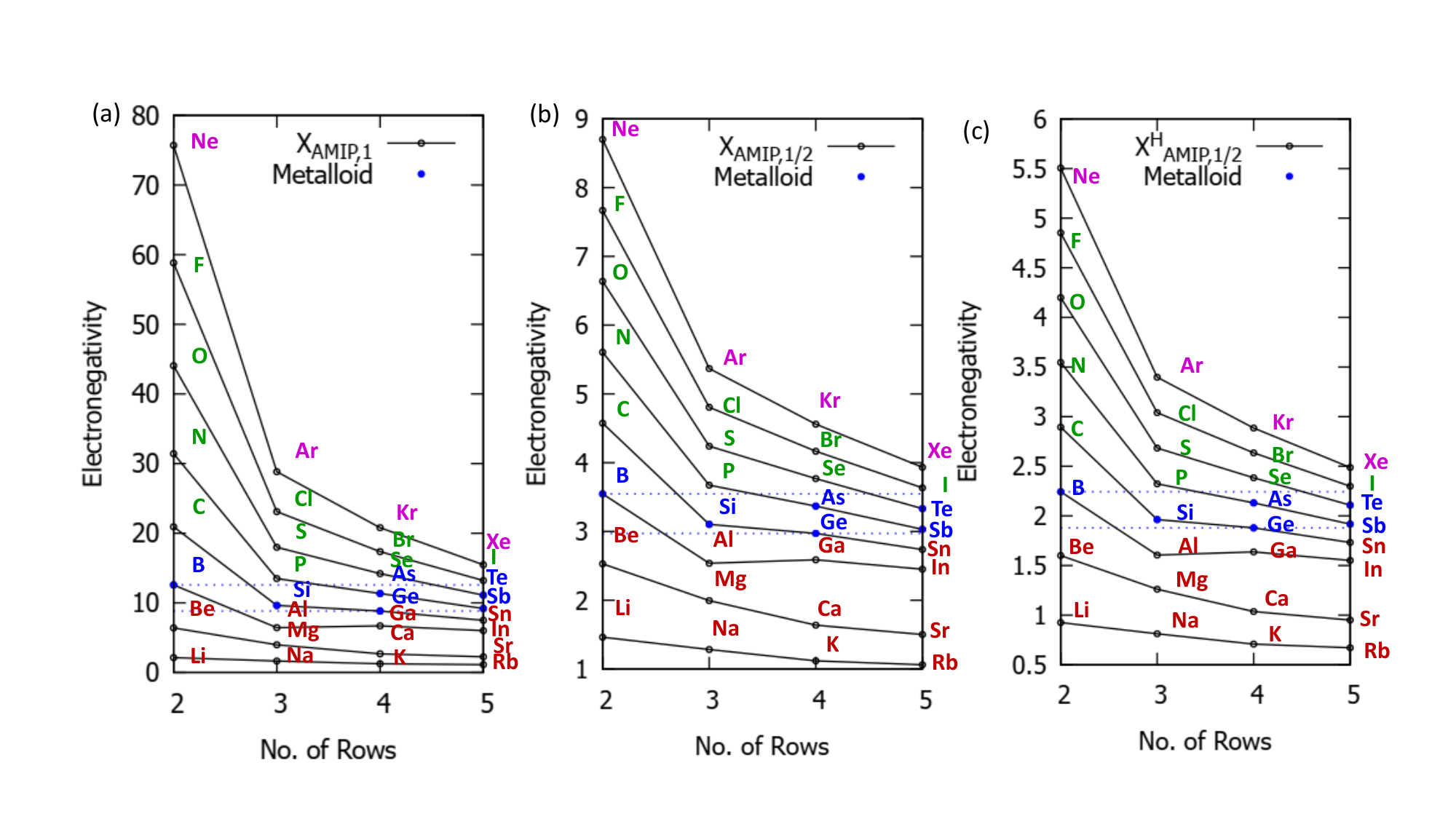}
 \caption{Metalloid-band (“Si-rule”) validation of the proposed electronegativity scales.  
Panels (a)–(c) plot the electronegativity of main-group elements versus the periodic-table row number for  
(a) $\chi_{\mathrm{AMIP},1}$,  
(b) $\chi_{\mathrm{AMIP},1/2}$, and  
(c) $\chi_{\mathrm{AMIP},1/2}^{H}$ (hydrogen-normalized).  
Open black circles connected by lines trace the periodic trends for each group.  
Elements traditionally classified as metalloids—B, Si, Ge, As, Sb, and Te—are highlighted by blue symbols and dotted horizontal guides, showing that they fall within a distinct metalloid band for all three AMIP-based scales.  
Nonmetals (green), noble gases (magenta), and metals (red) are labeled for clarity.  
The consistent placement of the metalloid band demonstrates that the AMIP electronegativity formalism accurately separates metals, metalloids, and nonmetals without misclassification.
}
  \label{fig:Fig6_metalloid_XAMIP}
\end{figure}

The canonical metalloids—boron (B), silicon (Si), germanium (Ge), arsenic (As), antimony (Sb), and tellurium (Te)—are highlighted in Fig. \ref{fig:Fig6_metalloid_XAMIP}. The position of this ``metalloid band" is highly sensitive to the accuracy of electronegativity values, providing a stringent test for any proposed scale~\cite{Allen1994}.

As shown in Fig. \ref{fig:Fig6_metalloid_XAMIP}(a)–(c), our scales—$\chi_{\mathrm{AMIP},1}$, $\chi_{\mathrm{AMIP},1/2}$, and $\chi_{\mathrm{AMIP},1/2}^H$—correctly place these elements within the metalloid band without any misclassifications. This performance matches that of the classic scales by Pauling and Allen. The agreement is particularly notable because our approach derives this classification using only three fundamental quantities: the principal quantum number, the atomic valence radius, and the zero-angle electron-scattering factor. From these parameters alone, the metalloid band emerges naturally, cleanly separating metals from nonmetals without any empirical adjustments.

Critically, all three AMIP-based expressions yield a consistent and physically intuitive classification: elements with electronegativities lower than the metalloid band (toward the left of the periodic table) are identified as metals, while those with higher values (to the right) are classified as nonmetals.

The capacity of electronegativity to delineate metals from nonmetals is a long-established concept. Pauling identified this distinction as the most obvious correlation between electronegativity and elemental behavior, a point later reinforced by Allen, who emphasized the metalloid band as a rigorous benchmark for validating new scales~\cite{Allen1994}. The ability of our mean-inner-potential–based scales to meet this benchmark underscores their physical soundness and broad applicability.

\subsection{Bonding Classification via Electronegativity}

\begin{figure}
\centering
\includegraphics[width=0.7\textwidth, angle=-90]{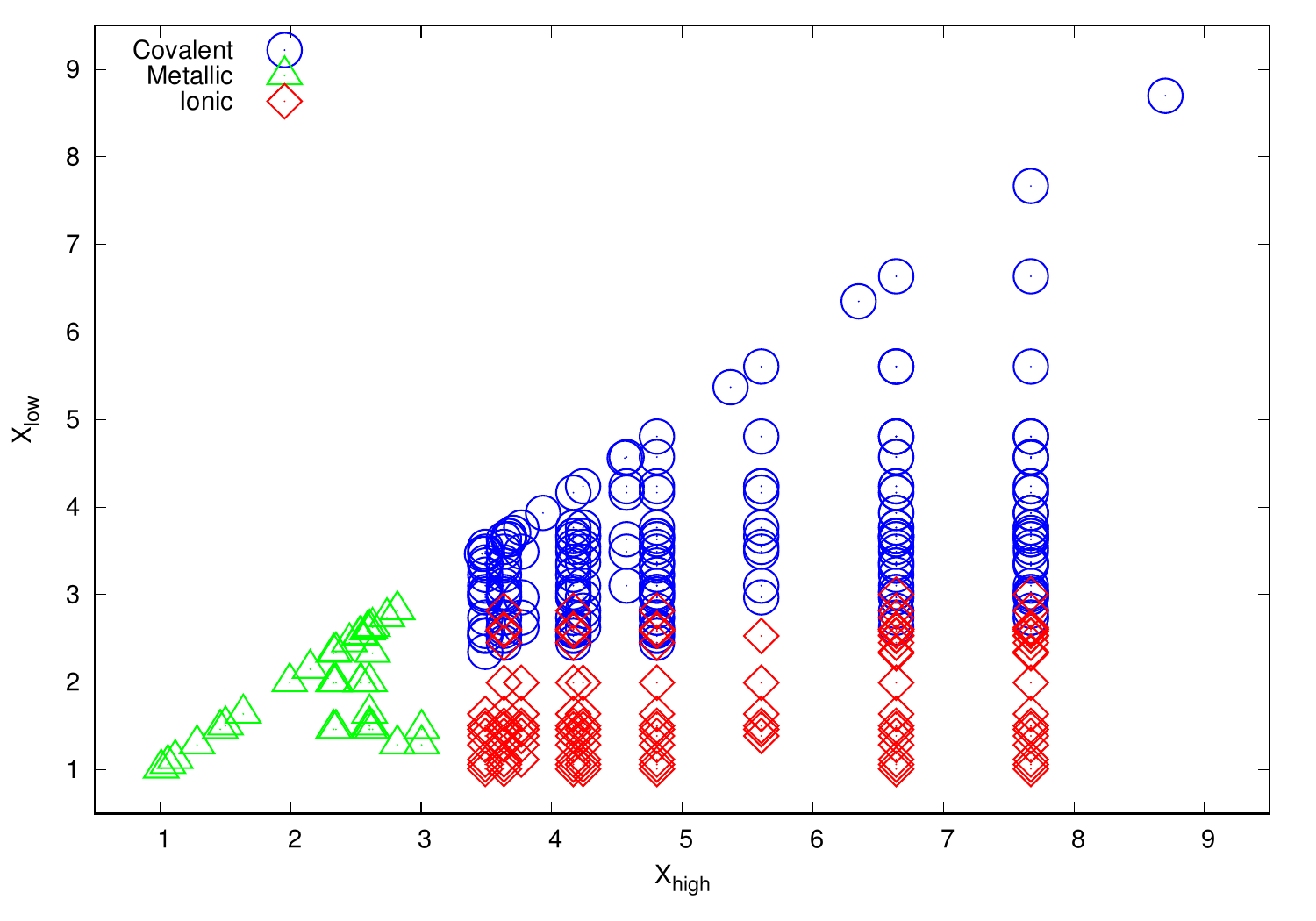}
\caption{Bonding classification of 358 compounds~\cite{meek2005electronegativity} using our proposed electronegativity scale $\chi_{\mathrm{AMIP},1/2}$. Compounds are classified as covalent (109, circles), metallic (36, triangles), or ionic (213, diamonds) based on the electronegativity values of their constituent elements. Each compound is plotted according to the higher ($\chi_{\mathrm{high}}$) and lower ($\chi_{\mathrm{low}}$) electronegativity values of its elements. The distinct regions occupied by each bond type illustrate the utility of electronegativity differences in predicting the primary bonding character of compounds.}
\label{fig:Fig7_Bond-triangle}
\end{figure}

A triangular bonding diagram~\cite{meek2005electronegativity,sproul2020evaluation,sproul2021cardinal}, constructed using our electronegativity scale ($\chi_{\mathrm{AMIP},1/2}$) for 358 compounds (Fig. \ref{fig:Fig7_Bond-triangle}), reveals a clear separation into three distinct regions corresponding to metallic (M), ionic (I), and covalent (C) bonding types. A pronounced gap along the $\chi_{\mathrm{hi}}$ axis cleanly separates the metallic region from the combined ionic/covalent region, while the overlap between ionic and covalent compounds along the $\chi_{\mathrm{lo}}$ axis is narrow. These characteristics align with the key indicators of a high-quality electronegativity scale as established by Sproul~\cite{sproul2020evaluation}.

Applying Sproul’s evaluation procedure~\cite{sproul2020evaluation}, our scale achieves a quality score of approximately 6, ranking it among the top-performing historical and modern electronegativity scales. This score is comparable to those of the best-performing scales, such as those by Allen~\cite{Allen1989}, Martynov–Batsanov~\cite{martynov1980new}, and Nagle~\cite{nagle1990atomic}, and is markedly superior to widely used but less discriminating scales like Pauling’s, which typically yield scores of 9–10~\cite{sproul2020evaluation}. The minimal ionic–covalent overlap indicates that very few binary compounds are misclassified relative to their accepted bonding character, while the significant metallic–nonmetallic gap reflects an accurate upper limit for metallic electronegativities.

These results demonstrate that our scale effectively captures the fundamental electronic axes described by Sproul~\cite{sproul2020evaluation}: $\chi_{\mathrm{lo}}$ correlates with bond polarity, while $\chi_{\mathrm{hi}}$ correlates with electronic conduction. The scale rivals or surpasses the best historical scales in distinguishing bonding types, providing independent confirmation that our electronegativity definition embodies the essential physics of chemical bonding. Consequently, it serves as a robust reference for future materials design and theoretical analysis.

\subsection{Correlation with Lewis Acid Strength}

\begin{figure}
\centering
\includegraphics[width=0.7\textwidth, angle=-90]{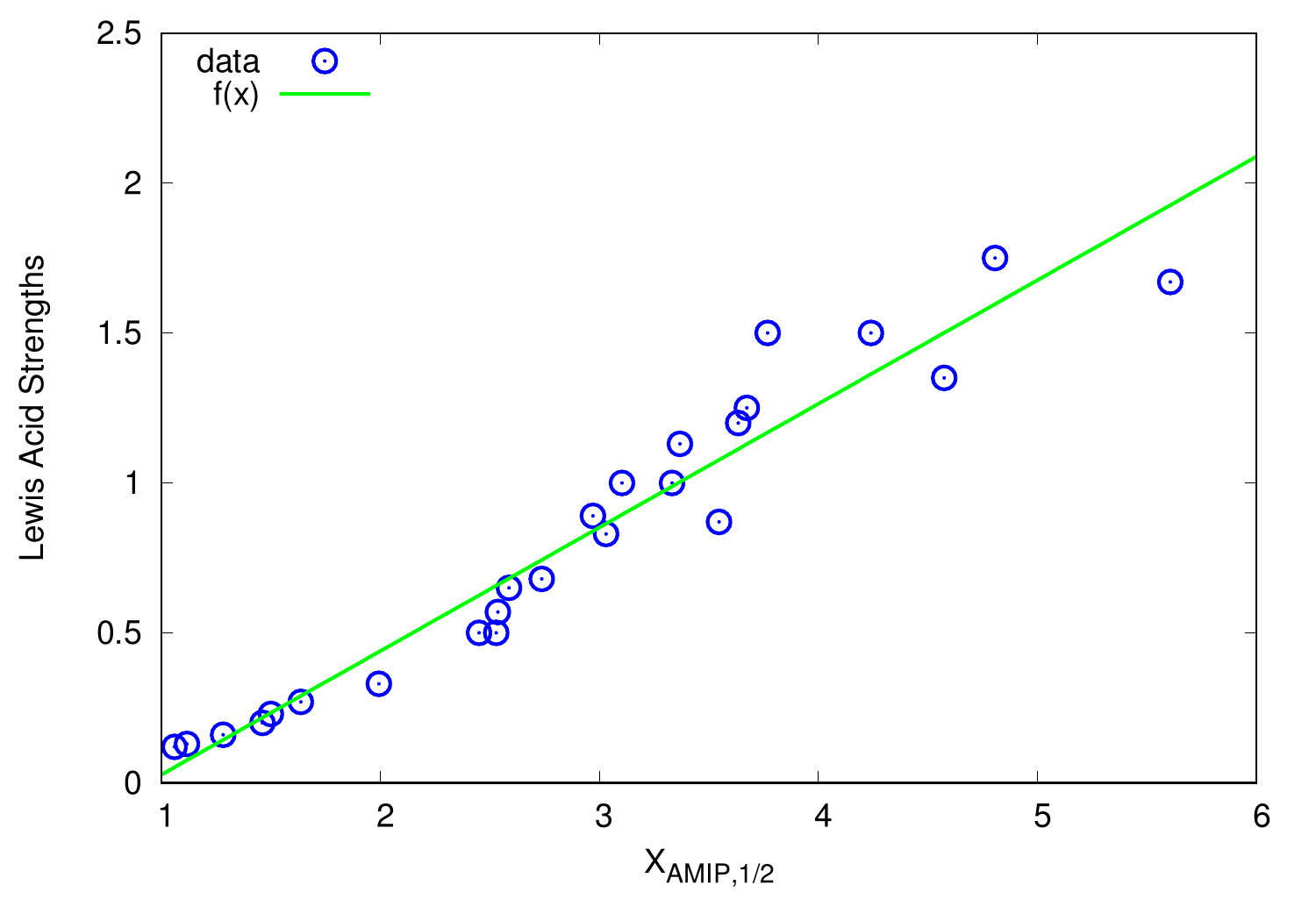}
\caption{Correlation between Lewis acid strength and the proposed electronegativity scale $\chi_{\mathrm{AMIP},1/2}$. Data comprise more than 14,000 experimentally characterized coordination environments across nearly 100 distinct cations~\cite{brown1988factors}. The best-fit line ($f(x)=0.4123x-0.3852$) shows an excellent linear relationship ($R^{2}=0.9347$), demonstrating that $\chi_{\mathrm{AMIP},1/2}$ accurately predicts Lewis acid strength across a broad chemical space.}
\label{fig:Fig8_Lewis-Acid}
\end{figure}

A strong linear correlation between Lewis acid strength~\cite{brown1988factors} and our electronegativity scale, $\chi_{\mathrm{AMIP},1/2}$, underscores the predictive power of this physically grounded descriptor. This analysis draws upon a comprehensive dataset of over 14,000 coordination environments encompassing nearly 100 distinct cations~\cite{brown1988factors}, providing a rigorous test of Lewis acidity (see Fig. \ref{fig:Fig8_Lewis-Acid}). The resulting coefficient of determination ($R^{2}=0.9347$) indicates that $\chi_{\mathrm{AMIP},1/2}$ effectively captures the fundamental electronic factors governing a cation’s ability to accept electron density from donor ligands.

This finding holds significant implications for both fundamental chemistry and materials science. Lewis acidity plays a critical role in diverse processes, including catalytic activity, solid-state ion transport, acid–base reactions, and the stability of coordination compounds, making a reliable quantitative predictor highly valuable. Derived from the atomic mean inner potential and requiring only elemental information, $\chi_{\mathrm{AMIP},1/2}$ offers a transferable, first-principles approach to estimating Lewis acidity across a wide range of chemical systems. The large-scale validation presented here demonstrates that our scale is not only a high-quality electronegativity measure according to Sproul’s bonding-separation criteria, but also a robust tool for predicting and designing materials where Lewis acid strength is a key functional parameter.

\subsection{Correlation with $\gamma$-Ray Spectral Width}

\begin{figure}
  \centering
  \includegraphics[width=0.7\textwidth, angle=-90]{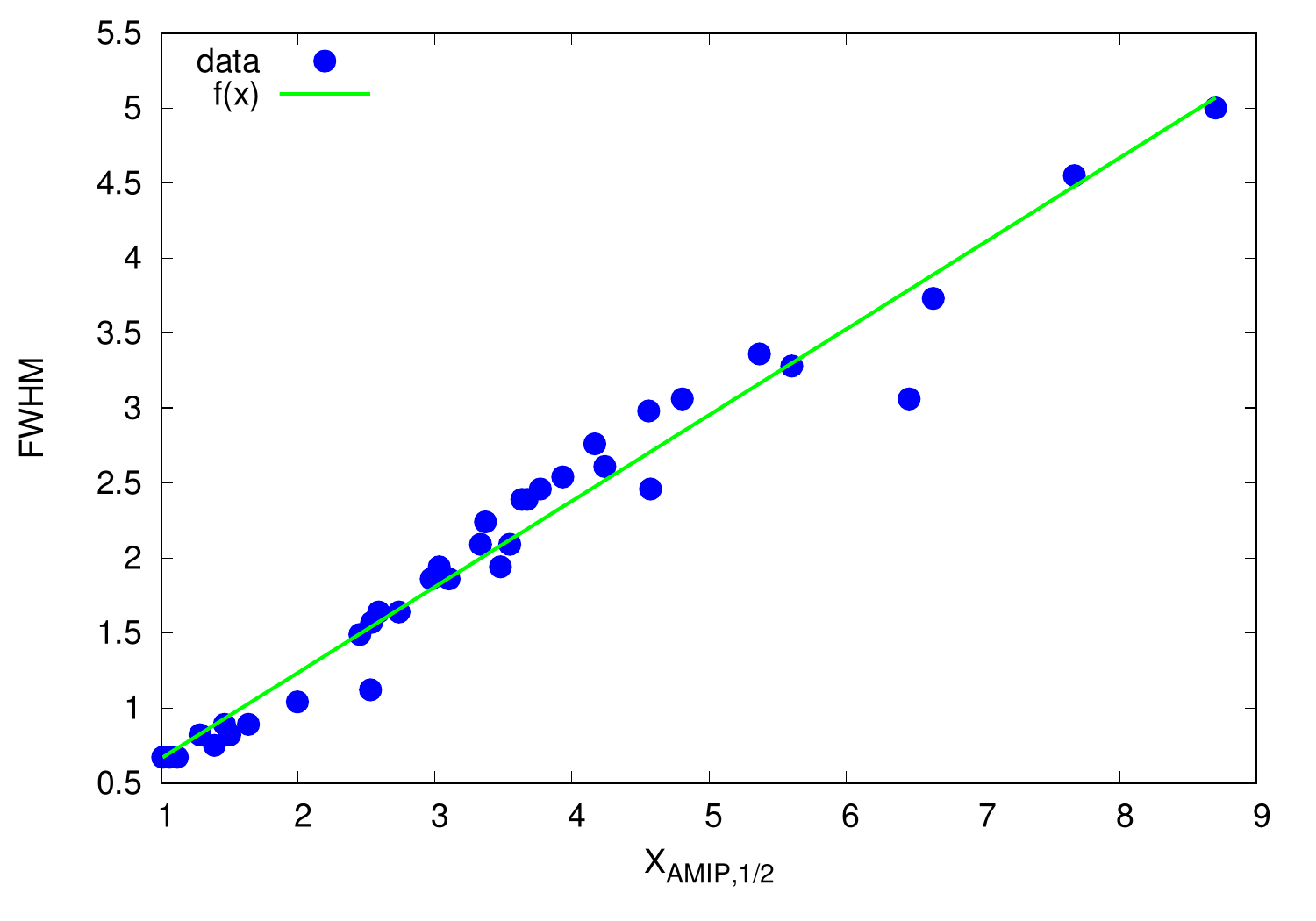}
\caption{Linear correlation between the full width at half maximum (FWHM) of the calculated $\gamma$-ray spectra and the proposed electronegativity scale $\chi_{\mathrm{AMIP},1/2}$.  
Data include the first 36 representative elements (Groups 1, 2, and 13–18).  
The best-fit line $f(x)=0.572\,x+0.0916$ (solid) yields $R^{2}=0.9656$, exceeding the reported correlation for Rahm’s electronegativity scale ($R^{2}\approx0.93$), demonstrating the strong predictive ability of $\chi_{\mathrm{AMIP},1/2}$ for $\gamma$-ray spectral widths.}
  \label{fig:Fig9_gamma-ray}
\end{figure}

The full width at half maximum (FWHM) of positron–electron annihilation $\gamma$-ray spectra reflects the momentum distribution of annihilating electron–positron pairs and, consequently, the underlying electronic structure of the atom. Previous work demonstrated a linear correlation between FWHM and Rahm’s electronegativity scale ($R^{2} \approx 0.93$), suggesting that atoms with a greater tendency to attract electrons yield broader $\gamma$-ray spectral widths~\cite{wu2021alternative}.

Our analysis reveals an even stronger correlation when FWHM is plotted against our electronegativity scale, $\chi_{\mathrm{AMIP},1/2}$ (See Fig. ~\ref{fig:Fig9_gamma-ray}). A least-squares fit yields:
\begin{equation}
\mathrm{FWHM} = 0.572 \chi_{\mathrm{AMIP},1/2} + 0.0916,
\end{equation}
with an excellent coefficient of determination $R^{2} = 0.9656$.

This enhanced correlation indicates that $\chi_{\mathrm{AMIP},1/2}$ more accurately captures the key physical factors—mean inner potential, charge density, and valence-electron extent—that govern the momentum distribution of positrophilic electrons. Since $\gamma$-ray annihilation spectroscopy probes valence electron density directly, the strong linearity underscores that our scale provides a quantitative link between atomic electronic structure and measurable high-energy observables. This predictive capability is valuable not only for interpreting positron annihilation experiments but also for modeling electron–positron interactions in condensed-matter systems and guiding the design of materials where precise knowledge of electron density distributions is critical.

\section{Discussions}

\subsection{Dependence of the AMIP-Based Electronegativity on the Exchange--Correlation Functional}

While the revPBE exchange--correlation functional was adopted in the present work owing to its well-established performance in describing structural and electronic properties, it is important to assess the sensitivity of the results to the choice of functional. To this end, Table~\ref{tab:chi_amip_light} compares the dimensionless electronegativity values derived from the proposed $\chi^{H}_{\mathrm{AMIP},1/2}$ scale for representative light elements (H--Ne), calculated using the LDA~\cite{perdew1981self}, PBE~\cite{perdew1996generalized}, and revPBE~\cite{zhang1998comment} functionals.

As shown in the table, all three exchange--correlation functionals yield qualitatively consistent trends in both the atomic mean inner potential and the corresponding electronegativity values, with only minor quantitative differences. The maximum deviations in $\chi^{H}_{\mathrm{AMIP},1/2}$ obtained using different functionals are small: approximately 0.020 between revPBE and LDA, 0.014 between PBE and LDA, and 0.006 between revPBE and PBE, following the systematic sequence LDA $>$ PBE $>$ revPBE. These variations are modest, indicating that the proposed electronegativity scale is robust with respect to the choice of exchange--correlation functional.

Importantly, the relative ordering of elements is fully preserved across all functionals considered, supporting the reliability, transferability, and general applicability of the AMIP-based electronegativity scale.

\begin{table}[htbp]
    \centering
    \caption{Calculated dimensionless electronegativity values based on the proposed $\chi^{H}_{\mathrm{AMIP},1/2}$ scale for elements from H to Ne, obtained using different exchange--correlation functionals: LDA, PBE, and revPBE.}
    \label{tab:chi_amip_light}
    \begin{ruledtabular}
        \begin{tabular}{ccccc}
            $Z$ & Element & LDA & PBE & revPBE \\
            \hline
            1  & H  & 2.200 & 2.200 & 2.200 \\
            2  & He & 4.093 & 4.090 & 4.088 \\
            3  & Li & 0.938 & 0.931 & 0.925 \\
            4  & Be & 1.616 & 1.607 & 1.600 \\
            5  & B  & 2.261 & 2.250 & 2.243 \\
            6  & C  & 2.913 & 2.900 & 2.894 \\
            7  & N  & 3.566 & 3.552 & 3.546 \\
            8  & O  & 4.219 & 4.205 & 4.199 \\
            9  & F  & 4.872 & 4.858 & 4.852 \\
            10 & Ne & 5.524 & 5.510 & 5.504 \\
        \end{tabular}
    \end{ruledtabular}
\end{table}

\subsection{Relation to Electronegativity as Electronic Chemical Potential}
\label{subsec:relation_to_chemical_potential}

In revising this work, we have sought to more accurately position the proposed AMIP-based electronegativity scale within the broader theoretical landscape. A key aspect of this contextualization involves clarifying its relationship to the widely recognized definitions of electronegativity rooted in the concept of the electronic chemical potential. Historically, the Mulliken scale established a foundational and physically profound link by defining electronegativity as the negative of the chemical potential for an atom. This concept was later rigorously formalized within modern density functional theory (DFT) by Parr and co-workers~\cite{Parr1978}, where electronegativity is identified with the negative of the electronic chemical potential:

\begin{equation}
    \chi = -\mu = -\left( \frac{\partial E}{\partial N} \right)_{v(\mathbf{r})},
\end{equation}

typically evaluated via finite differences of the total energy with respect to electron number. These frameworks establish a deep and general quantum-mechanical basis for electronegativity, directly connecting it to the energetic response of a system.

Our present approach, while also a first-principles construct, derives electronegativity from a different fundamental property: the atomic mean inner potential (AMIP). The AMIP characterizes the spherically averaged electrostatic potential generated by the atomic charge density. Consequently, it probes a complementary physical aspect---the internal electrostatic environment---rather than the response of the total energy to changes in electron number. This distinction clarifies that the AMIP-based scale is not intended to supplant these well-established chemical potential-based definitions. Instead, it offers an alternative and complementary perspective for quantifying electronegativity, one that is directly accessible from ground-state electronic structure calculations without requiring finite-difference energy evaluations. Both paradigms, stemming from different vantage points within electronic structure theory, serve to enrich the understanding of this central chemical concept.

\subsection{Experimental accessibility and uncertainty of AMIP}
\label{subsec:amip_experiment}

The atomic mean inner potential (AMIP) introduced in this work is closely connected to experimentally measurable quantities obtained from electron-based techniques, particularly electron holography and electron scattering. In practice, such experiments probe the \emph{mean inner potential} (MIP) of crystalline or amorphous solids rather than isolated free atoms. The experimentally determined quantity therefore corresponds to a crystalline mean inner potential $V_0$, from which the zero-angle electron scattering factor $f^{(e)}(0)$ can be extracted through well-established theoretical relations. Once $f^{(e)}(0)$ is known, an atomic-level AMIP can be obtained by introducing a well-defined atomic volume $\Omega$, according to
\begin{equation}
v_0 = K\,\frac{f^{(e)}(0)}{\Omega},
\end{equation}
where $K$ is a known proportionality constant. The choice of $\Omega$ thus plays a central role in connecting experimentally accessible crystalline quantities to atomic-scale descriptors.

The experimental feasibility and associated uncertainty of this procedure can be illustrated using silicon as a representative example. Electron holography measurements of the MIP of crystalline Si reported in the literature span a range of values. Early off-axis electron holography combined with theoretical corrections~\cite{gajdardziska1993accurate} yielded $V_0 = 9.26 \pm 0.08$~V, while subsequent holographic studies reported values~\cite{wang1997transmission,wu2004unique,kruse2006determination} between approximately $11.5$ and $12.5$~V, with uncertainties ranging from a few tenths of an electron volt up to about $1.3$~V. These variations reflect differences in experimental conditions, specimen thickness determination, treatment of dynamical diffraction effects, and phase-noise limitations inherent to electron holography.

Using the established relationship between $V_0$ and the forward electron scattering factor $f^{(e)}(0)$, these experimental MIP values correspond to a spread in $f^{(e)}(0)$ of roughly $3.9$--$5.3$~\AA{} for crystalline Si, with the largest uncertainty on the order of $0.5$~\AA{}. By contrast, our first-principles calculations for the Si pro-crystal yield $V_0 = 13.70$~V and $f^{(e)}(0) = 5.735$~\AA{}, in excellent agreement with independent Hartree--Fock calculations ($V_0 = 13.91$~V)~\cite{kruse2006determination}. The systematically smaller experimental values observed for crystalline Si are expected due to bonding-induced charge redistribution relative to a superposition of neutral atoms.

For crystalline materials, once $V_0$ or $f^{(e)}(0)$ has been experimentally determined, a crystalline AMIP-based electronegativity can be defined provided that a consistent atomic volume partitioning scheme is adopted (e.g., based on crystallographic volumes). We emphasize that this volume definition constitutes a key calibration step in bridging experimental measurements and atomic-scale AMIP values. While different choices of $\Omega$ may lead to quantitative shifts, the physically meaningful trends remain robust when a consistent scheme is employed.

Overall, these considerations demonstrate that AMIP is not only a well-defined theoretical quantity but also one that is, in principle, experimentally accessible. The dominant sources of uncertainty---including specimen thickness calibration, phase noise, bonding effects, and atomic volume definition---are well understood within the electron microscopy and scattering communities. This establishes a realistic and physically transparent pathway for connecting electron-based measurements to the AMIP-derived electronegativity scale proposed in this work.

\subsection{Prospects for Pressure-Dependent Electronegativity}
\label{subsec:pressure}

The proposed electronegativity scale, being derived from the atomic mean inner potential (AMIP), is intrinsically linked to the electronic charge density. Because the charge density responds sensitively to external pressure---primarily through atomic volume compression and associated reorganization of the electronic structure---the present formalism is, in principle, naturally extensible to high-pressure environments. Under applied pressure, modifications of the charge density directly alter the AMIP and, consequently, the derived electronegativity values. A pressure-dependent electronegativity scale could therefore be constructed by evaluating AMIP from first-principles charge densities obtained at finite pressures.

The feasibility of pressure-dependent electronegativity has been demonstrated in several recent studies. Rahm \textit{et al.} showed that compression leads to systematic changes in atomic electron configurations and electronegativity across the periodic table, driven by density redistribution and orbital reordering under pressure~\cite{rahm2019squeezing}. More recently, Dong \textit{et al.} developed a pressure-dependent electronegativity and chemical hardness framework based on first-principles calculations, explicitly quantifying how electronic descriptors evolve with increasing pressure~\cite{dong2022electronegativity}. These works provide strong precedent that electronegativity, when formulated in terms of fundamental electronic quantities, can be meaningfully extended to extreme thermodynamic conditions.

Within the AMIP-based framework, pressure effects enter naturally through the pressure dependence of both the charge density and the effective atomic volume. In practice, a pressure-dependent AMIP could be obtained by computing the forward electron scattering factor from compressed charge densities and combining it with a pressure-dependent atomic volume derived from first-principles structural data. However, a consistent and transferable implementation of this approach requires careful methodological choices, particularly concerning the partitioning of charge density and the definition of atomic volumes under compression. These issues are nontrivial when aiming for a universal atomic-scale reference and are therefore beyond the scope of the present foundational study.

A systematic investigation of these methodological aspects, together with quantitative benchmarking against existing pressure-dependent electronegativity models, constitutes a clear and promising direction for future work. Such an extension would enable the present AMIP-based electronegativity scale to be applied to materials and chemical systems under extreme conditions, further broadening its applicability.

\subsection{Limitations}

We also note several limitations of the present AMIP-based electronegativity approach. The method is formulated within a ground-state, first-principles framework and is therefore most straightforwardly applied to closed-shell atoms, where the electronic density is well defined. However, the approach can also be applied to open-shell systems, provided that the electronic charge density is calculated accurately, allowing proper treatment of spin polarization and correlation effects. These considerations do not affect the main conclusions of the present work for the elements considered here, but they highlight important directions for future development aimed at extending and validating the AMIP-based electronegativity scale for more complex electronic configurations.

\section{Conclusion}

We present a family of electronegativity scales derived directly from the atomic mean inner potential (AMIP), 
$\chi_{\text{AMIP},p} = \left(\frac{v_0}{n_q}\right)^{p}
= \left[K\frac{f^{(e)}(0)}{n_q \Omega_a}\right]^{p} = \left[ \frac{3K}{4\pi}
\frac{f^{(e)}(0)}{n_q (r_v)^3}\right]^{p} = \left[ \frac{K}{4\pi}
\frac{\langle r_t^2 \rangle}{n_q (r_v)^3}\right]^{p}$, 
with $p = 1$ or $1/2$. 
These scales require only three ground-state quantities for each element: 
the principal quantum number ($n_q$), the forward electron-scattering factor $f^{(e)}(0)$, and the atomic volume ($\Omega_a$). 
Equivalently, the latter two quantities are related to the second moment of the total charge density ($\langle r_t^2 \rangle$) 
and the first moment of the valence charge density ($r_v$).

The linear form $\chi_{\mathrm{AMIP},1}$ and its square-root variant $\chi_{\mathrm{AMIP},1/2}$ emerge analytically without empirical parameters and possess clear physical units.  
A dimensionless Pauling-referenced scale, $\chi_{\mathrm{AMIP},1/2}^{H}$, is obtained by simple normalization to hydrogen.

Benchmarking shows that $\chi_{\mathrm{AMIP},1/2}$ closely tracks established electronegativity definitions, with coefficients of determination $R^{2}>0.87$ for main-group elements and $R^{2}>0.80$ for whole periodic table including transition metals and anthanides and actinides. All AMIP-based expressions correctly position the canonical metalloids (B, Si, Ge, As, Sb, Te) on the metal–nonmetal boundary, matching the performance of the best historical scales.

Because every term in the AMIP formalism is a well-defined quantum-mechanical observable, the method is parameter-free, readily implemented in first-principles calculations, and computationally efficient.  
Its physical transparency, strong agreement with both classical and modern reference scales, and accurate treatment of the metal–nonmetal divide establish AMIP electronegativity as a concise and predictive descriptor of chemical bonding and reactivity across the periodic table.

We also showcase applications of our proposed AMIP-based electronegativity, including the classification of metals and nonmetals, bonding classification based on electronegativity, correlations with Lewis acid strength, and correlations with $\gamma$-ray spectral width. Recently, we successfully established an electronegativity-based descriptor for hydrogen adsorption on group IV and III--V two-dimensional honeycomb structures~\cite{ng2025partitioning} using our previously proposed scattering electronegativity~\cite{zheng2024universal}. Given the high coefficient of determination ($R^2 > 0.90$) between the scattering electronegativity and the AMIP-based electronegativity, further promising applications of AMIP-based electronegativity in energy research are expected.

\begin{acknowledgments}
This research was supported by the National Natural Science Foundation of
China (NSFC) (No. 12274355) and the Xiamen University Malaysia Research Fund (No. XMUMRF/2024-C14/IORI/0005).
\end{acknowledgments}

\bibliography{EN-refs}

\end{document}